# Temperature Profiles of Hot Gas In Early Type Galaxies

Dong-Woo Kim[1], Liam Traynor[1,2], Alessandro Paggi[1,3,4], Ewan O'Sullivan[1], Craig Anderson[1], Douglas Burke[1], Raffaele D'Abrusco[1], Giuseppina Fabbiano[1], Antonella Fruscione[1], Jennifer Lauer[1], Michael McCollough[1], Douglas Morgan[1], Amy Mossman[1], Saeqa Vrtilek[1], and Ginevra Trinchieri[5]

[1] Harvard-Smithsonian Center for Astrophysics, 60 Garden Street, Cambridge, MA 02138, USA
[2] University of Southampton
[3] INAF-Osservatorio Astrofisico di Torino, via Osservatorio 20, 10025 Pino Torinese, Italy
[4] INFN–Istituto Nazionale di Fisica Nucleare, Sezione di Torino, via Pietro Giuria 1, 10125 Turin, Italy
[5] INAF-Osservatorio Astronomico di Brera, Via Brera 28, I-20121 Milan, Italy

December 11, 2019

## ABSTRACT

Using the data products of the Chandra Galaxy Atlas (Kim et al. 2019a), we have investigated the radial profiles of the hot gas temperature in 60 early type galaxies. Considering the characteristic temperature and radius of the peak, dip, and break (when scaled by the gas temperature and virial radius of each galaxy), we propose a *universal* temperature profile of the hot halo in ETGs. In this scheme, the hot gas temperature peaks at $R_{MAX} = 35 \pm 25$ kpc (or ~0.04 $R_{VIR}$) and declines both inward and outward. The temperature dips (or breaks) at $R_{MIN}$ (or $R_{BREAK}$) = 3 - 5 kpc (or ~0.006 $R_{VIR}$). The mean slope between $R_{MIN}$ ($R_{BREAK}$) and $R_{MAX}$ is $0.3 \pm 0.1$. Allowing for selection effects and observational limits, we find that the universal temperature profile can describe the temperature profiles of 72% (possibly up to 82%) of our ETG sample. The remaining ETGs (18%) with irregular or monotonically declining profiles do not fit the universal profile and require another explanation. The temperature gradient inside $R_{MIN}$ ($R_{BREAK}$) varies widely, indicating different degrees of additional heating at small radii. Investigating the nature of the hot core (HC with a negative gradient inside $R_{MIN}$), we find that HC is most clearly visible in small galaxies. Searching for potential clues associated with stellar, AGN feedback, and gravitational heating, we find that HC may be related to recent star formation. But we see no clear evidence that AGN feedback and gravitational heating play any significant role for HC.

# 1. INTRODUCTION

The hot gaseous halos of early-type galaxies (ETGs) provide crucial information for the formation and evolution of the host galaxy. Various physical processes affecting the galaxy evolution are also reflected in the thermal structure of the hot ISM (e.g., see Kim & Fabbiano 2015). They include mergers, the infall of gas, ram pressure stripping, sloshing (and other tidal interactions), AGN feedback, and stellar feedback.

Early models for the hot halos of ETGs suggested cooling flows because the predicated cooling time is shorter than the Hubble time (e.g., Fabian 1994; Sarazin & White 1987). With a rapid radiative cooling in the central region, these models predict the temperature to rapidly decrease and the surface brightness to strongly peak towards the center. However, observations have shown that the large amount of expected cool gas is not present, and the cooling occurs on a much smaller scale than predicted (e.g., Fabian 2012 and references therein). This implies that there is a source of internal heating that prevents rapid cooling in the center of the hot halo. The possible heating mechanisms include AGN feedback (e.g., Fabian 2012), stellar feedback (e.g., Ciotti et al. 1991; Tang & Wang 2005), and gravitational heating (e.g., Khosroshahi et al. 2004; Johansson et al. 2009). Different models predict different temperature profiles. The classical cooling flow model predicts that the temperature peaks at a certain radius and declines both inward and outward (e.g., Sarazin & White 1987; Pellegrini 2012). Recent simulations of pure cooling flows indicate that the temperature declines toward the center, roughly following $T \sim r^{0.5}$ (Gaspari et al. 2012). In contrast, feedback models predict different degrees of temperature increase toward the center. Recent hydrodynamic simulations of AGN feedback (e.g., Pellegrini et al. 2012a, Ciotti et al. 2017 – they assume AGN winds, but no jet) show that the gas temperature monotonically decreases to the center from a few effective radii, but the profile becomes somewhat chaotic during the short period of major bursts. Possible heating mechanisms such as SN heating (e.g., Ciotti et al. 1991; Tang & Wang 2005), gravitational heating from the SMBH (Pellegrini et al. 2012b), and gravitational potential energy during infall (Khosroshahi et al. 2004) are also reflected in the profiles. These models provide an alternative to the cooling flow models. However, they still do not capture the full complexity of the temperature profiles for ETGs, and more comprehensive models are needed to be able to explain all the features seen in these temperature profiles. To advance our understanding of the various mechanisms and to help to leverage various model parameters, it is necessary to provide accurate observational constrains on the T profiles.

It has been suggested that galaxy groups and clusters may have a universal temperature profile (when the core is excluded) that is close to self-similar (e.g., De Grandi & Molendi 2002; Vikhlinin et al. 2005; Sanderson et al. 2006; Sun et al. 2009). In this picture, the temperature is rising rapidly with increasing radius out to r ~ 0.1 $R_{VIR}$ (viral radius), before slowly decreasing to large radii. The temperature of the gas is primarily governed by the gravity of the groups and clusters. In ETGs (and small groups), however, baryonic physics becomes more important, while the gravitational effects (self-similarity) become less dominant than clusters and large groups. The presence of X-ray cavities hollowed out by radio jets, SN driven galactic winds, filaments, shells, tidal features and cold fronts (e.g. Böhringer et al. 1993; Churazov et al. 2001; Colbert et al. 2001; Gastaldello et al. 2008) all impact the temperature profiles of ETGs which can vary widely from one galaxy to another. For example, some galaxies show a temperature decrease toward the center and peaks at large radii, while others show a rising temperature towards their center (e.g., Diehl & Statler 2008; Pellegrini et al. 2012b). See also O'Sullivan et al. (2017) that some groups also show the temperature rising toward the center.

In this paper, we use a sample of 60 nearby ETGs taken from the Chandra Galaxy Atlas (Kim et al. 2019a) with extended X-ray emission that allows us to extract temperature profiles. As described in Kim et al. (2019a), our sample includes several examples of brightest group/cluster galaxies (BCGs). However, we exclude large groups/clusters by limiting $T_{GAS}$ (determined from the entire hot halo) below ~1.5 keV, because $T_{GAS}$ is a good measure of the total mass of the system. About 80% of the sample galaxies have $T_{GAS}$ = 0.3–1.0 keV. For comparison, the previous detailed study of the temperature profiles by Diehl & Statler (2008) (DS08 hereafter) included the temperature profiles of 36 ETGs with four years of Chandra data. We investigate the temperature profiles of hot halos in great detail, both at large scales by examining global properties and profile trends as well as at small scales by examining the profiles within the central region of the galaxy. We compare features (e.g., peaks and dips) found in each profile to look for similarities in shape and to explore a possibility of the presence of a "universal" profile. We then test which galaxy properties play a role in the formation of these features.

In Section 2, we show the data reduction methods and how we determine the 3D temperature profile. In section 3, we describe six types of temperature profiles. In section 4, we describe the common characteristics, and in section 5, we explore the possibility of a universal temperature profile. In section 6, we further investigate the inner temperature profile and the correlation with other galaxy properties. Finally, we summarize our results in section 7.

## 2. DATA REDUCTION

### *2.1 Data Analysis*

The analysis of the archival Chandra data for the Chandra Galaxy Atlas (CGA) project is described in full by Kim et al. (2019a). In this paper, we will briefly describe the key steps used in this analysis. The initial step is to reprocess all Chandra data with a CIAO[1] tool ***chandra_repro***, to merge multiple observations, then exclude all point sources detected by a CIAO tool ***wavdetect.*** The point source size is determined by the point spread function (PSF)[2]. Because the PSF becomes large at large off-axis angles (OAA), we do not use observations where the target galaxy is at OAA> 4 arcmin. We remove the time intervals containing high background flares with the CIAO tool ***deflare***. Because each chip will be affected differently by background flares we apply this step per observation (specified by obsid) per chip (specified by ccdid).

This work makes use of the four adaptive spatial binning methods which were implemented in the CGA project to characterize the spectral properties of the hot gas; (1) annulus binning (AB) with adaptively determined inner and outer radii, (2) weighted Voronoi tessellation adaptive binning (WB; Diehl & Statler 2006), (3) contour binning (CB; Sanders 2006), and (4) hybrid binning (HB; O'Sullivan et al. 2014). AB provides azimuthally averaged quantities, and the latter three methods provide two-dimensional spectral properties. We examine all four binning results to identify the 2D thermal structure, but only use AB and WB for quantitative measurements and model fitting, because CB and HB produce radially overlapped regions. These binning methods are controlled by pre-set S/N. We use three S/N values during the spatial binning (20, 30, 50) to optimize the balance between resolution and statistics.

---

[1] http://cxc.harvard.edu/ciao/
[2] http://cxc.harvard.edu/ciao/PSFs/psf_central.html

Once the adaptive spatial binning is complete, the X-ray spectra are extracted from each spatial bin, per observation per chip. The corresponding arf and rmf files are also generated, per observation per chip, in order to take account of time- and position-dependent ACIS responses. To remove background emission, we download blank sky data from the Chandra archive; then, for each observation, we re-project them to the same tangent plane as done in the observations. We also rescale them to match the higher energy (9-12 keV) rate, where the photons are primarily from the background (Markevitch 2003). We compared our results with those made by the local background from the off-axis, source-free region from the same observation in a few cases, and found no significant difference.

After generating source spectrum, background spectrum, arf, and rmf per observation per chip, we use a CIAO tool, *combine_spectra*, to combine them to make a single data set per bin[3]. We also performed a joint fit by simultaneously fitting individual spectra and found no significant difference. We primarily use a two-component emission model, APEC, for hot gas and power-law for undetected LMXBs. We fix the power-law index to be 1.7, which is appropriate for the hard spectra of LMXBs (e.g., Boroson et al. 2011) and $N_H$ to be the Galactic HI column density (Dickey and Lockman 1990). We also fix the metal abundance to be solar at GRSA (Grevesse, N. & Sauval, A.J. 1998). Although the abundance is known to vary from a few tenth to a few times solar inside the hot ISM, the hot gas temperatures do not significantly depend on the abundance (e.g., see Kim & Pellegrini 2012).

*2.2 3D Temperature Profiles*

To adequately describe the 3D gas properties, we parameterize the 3D temperature and density profiles and find the best-fit parameters by projecting the 3D models and fitting them to the projected emissivity and projected temperature profiles. For temperature models, we use two models from Gastaldello et al. (2007), smoothly joined power laws and power laws mediated by an exponential. We also use one detailed by Vikhlinin et al. (2006), which has more parameters to fit, particularly for cooling cores and T peaks at $r \sim 0.1\ r_{200}$, often seen in typical clusters. With the temperature profiles showing a large variation in shape and structure, using all three models gives us the flexibility required to describe the various observed temperature profiles reasonably well. We apply two different initial values for each model before fitting to improve the ability of the models to converge fast for complex projected temperature profiles.

To compare with the projected T profile, we need to parameterize the density profile to calculate the emissivity. For density models, we use a single and double β model (e.g., Sarazin and Bahcall 1977) and one detailed by Vikhlinin et al. (2006), which has more parameters to fit, particularly for cooling cores and steep declining outskirts at large radii.

Since the gas temperature in ETGs is around 1 keV, the X-ray emission is dominated by emission lines over the thermal bremsstrahlung continuum. In this case, the temperature can be mainly determined by the energy at the peak intensity and this peak energy is linearly proportional to the plasma temperature in a log scale (see Figure 2 in Vikhlinin 2006). Following Vikhlinin (2006), we estimate a projected temperature and emissivity from each model in a given spatial bin and compare with observed values to determine the best-fit model parameters. For the profile fitting with Sherpa[4], we proceed as follows. For each 3D density model $n(r)$, we define a 2D projected surface density profile through the integral function[5]

$$S(R) = \int_{R}^{\infty} n(r)^2 \frac{r}{\sqrt{r^2 - R^2}} \, dr$$

and fit the observed surface brightness profiles. We compare our best-fit models using a $\chi^2$ method to select which model gives the best fit to the data. If multiple models are similarly acceptable, we select the simplest model with the smallest number of parameters. Then, we define the 2D projected temperature profile, $t(R)$, through the best-fit 3D density profile and the 3D temperature model, $T(r)$, using the integral function

$$\log t(R) = \frac{\int_{R}^{\infty} n(r)^2 \, \log T(r) \, \frac{r}{\sqrt{r^2 - R^2}} \, dr}{\int_{R}^{\infty} n(r)^2 \, \frac{r}{\sqrt{r^2 - R^2}} \, dr}$$

and fit the observed 2D projected temperature profiles. Again, we compare our best-fit models using a $\chi^2$ method to select which model gives the best fit to the data. We note that the density and temperature profiles parameterized by the corresponding models are useful, but not the models themselves as long as the models reproduce the observed profiles. The parameterized profiles can be used to measure the entropy, pressure and mass profiles and we will present them with proper abundance measurements in the next paper. In this paper, we focus on the shape of the temperature profile.

---

[3] This tool, *combine_spectra* with an option *method=sum*, makes the exposure time of the combined spectrum to be the sum of those of individual spectra. This is proper when multiple obsids are combined, but not when multiple chips (ccdid) of a given obsid are combined. In this case, we manually correct the exposure time to be the average value of multiple chips of the same obsid. Note that if *combine_spectra* is used with an option *method=avg* to get an average exposure time, the combined arf is the sum of individual files which is not applicable.

## 3. TEMPERATURE PROFILE TYPES

The isothermal hot ISM may be the simplest case, but we do not see such a case in a single galaxy. Instead, we always find radial variations with positive or negative gradients. Some galaxies have monotonically decreasing or increasing temperature profiles, while others show one or more breaks with bumps and/or dips in their temperature profiles. We categorize the observed temperature profiles in our sample into six types; hybrid-bump (rising at small radii and falling at large radii), hybrid-dip (falling at small radii and rising at large radii), double-break (falling at small radii, rising at intermediate radii, and falling again at large radii), positive (rising all the way), negative (falling all the way) and irregular. DS08 adopted four different types (hybrid-bump, positive, negative, quasi-isothermal). We have added three new types (hybrid-dip, double-break, and irregular), but excluded the isothermal type because we found no obvious isothermal case. We show an example of each profile type in Figure 1. For the entire sample of 60 ETGs, we present the observed temperature profiles and the best-fit models in Appendix A.

---

[4] http://cxc.harvard.edu/sherpa/
[5] Integrals are performed making use of the QUAD function from the python package, scipy.integrate.

To determine the type, we have examined the temperature profiles made in 4 different spatial binning methods and three different s/n values and corresponding 2D temperature maps. In particular, the 2D binning methods retain the spatial information, which provides 2D gas distribution and its temperature. This allows gas structure and/or any asymmetry in the halos to be identified. Non-spherically symmetric gas distributions manifest in the projected temperature profiles as vertical (temperature) scatter, due to the range of gas temperatures at the same radii. In this work, we intend to determine the shape of the global temperature profile and use the azimuthally averaged radial profiles. The azimuthal variation (e.g., analyzing different pie sectors) will be addressed in future work. If X-ray bright nearby galaxies exist within the X-ray halo of the target galaxy, we exclude the corresponding 2D spatial bins from the profile before fitting.

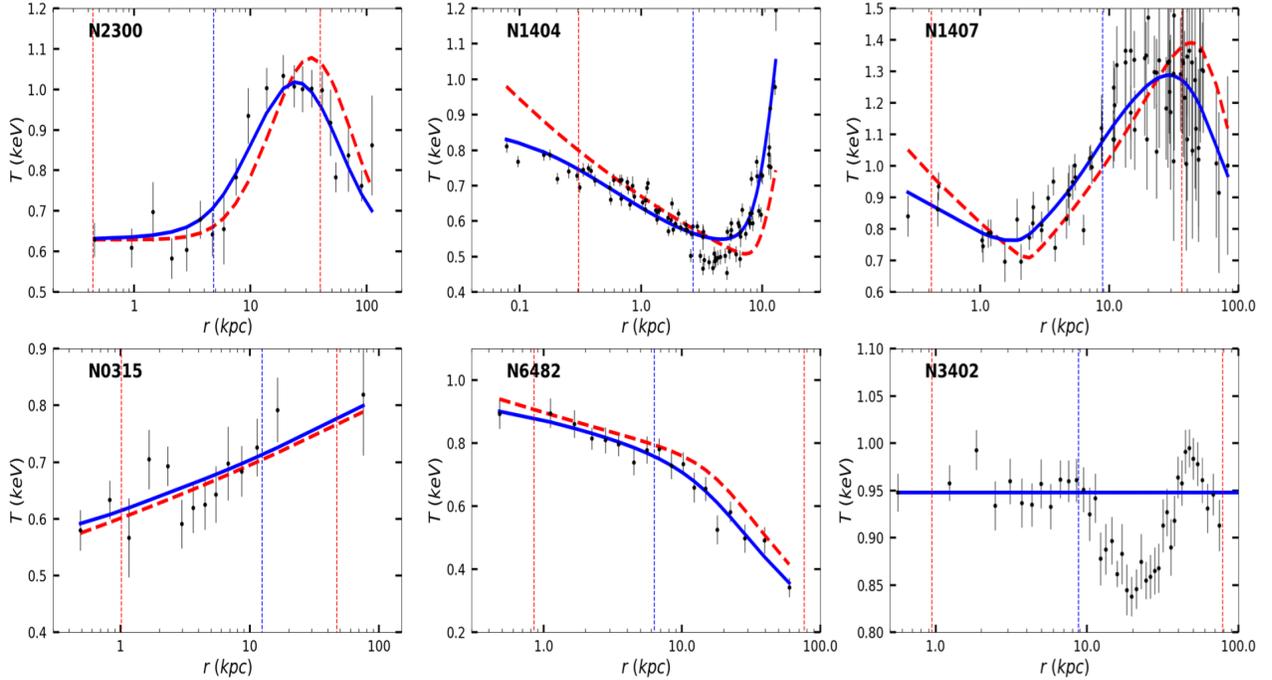

Figure 1. Examples of six different temperature profile types. All the galaxies in our sample are shown in Appendix A. From left to right and top to bottom; hybrid-bump (rising at small radii and falling at large radii), hybrid-dip (falling at small radii and rising at large radii), double-break (a profile containing both a dip and peak), positive (monotonically rising), negative (monotonically falling) and irregular. The data points in black are fitted with projected temperature profiles in blue with the 3D-model shown in a red dashed line. The inner red vertical line indicates r = 3" where the AGN could affect the temperature measurement, and the outer red line indicates the maximum radius where the hot gas emission is reliably detected with an azimuthal coverage larger than 95%. The blue vertical line is at one effective radius.

### 3.1 Hybrid-Bump

The temperature profiles in the hybrid-bump type all have a temperature peak in the middle of the observable radius range. The temperature gradient is positive inside the peak and negative outside the peak. This is most common, with 26 (or 43 %) out of 60 belonging to this type. In our sample, the peak with $T_{MAX}$ = 1-2 keV lies at $R_{MAX}$ = 10 – 70 kpc, or a few hundredth $R_{VIR}$ (see Table 2 for $T_{MAX}$ and $R_{MAX}$ and section 4 and 5 for more discussions). This type is similar to a typical profile of groups and clusters, except that $R_{MAX}$ is smaller than that found for clusters where the peak is at r ~ 0.1 $R_{VIR}$, or r ~100-200 kpc (for systems with T = 2- 10 keV).

The profiles in this type can be separated further into two subsamples based on their inner slope change. We find that 17 out of the 26 hybrid-bump profiles show a flattening at small radii. We compare two examples in Figure 2. In NGC 5129 (left panel), the temperature profiles show a constant gradient at small radii (r < $R_{MAX}$). In NGC 533 (right panel), the temperature increases slowly at small radii, rapidly increases at intermediate radii, and then decreases at large radii. The galacto-centric distance of the inner break is at $R_{BREAK}$ ~ 5 ± 2 kpc (see Table 2 and section 4 and 5). For a few distant galaxies (e.g., NGC 4104 at 120 Mpc), this radius falls within a few arcsec from the center such that the inner break would not be properly recognized due to the possible contamination of AGN. Therefore, the presence of the inner break may be more frequent than we could identify in this work.

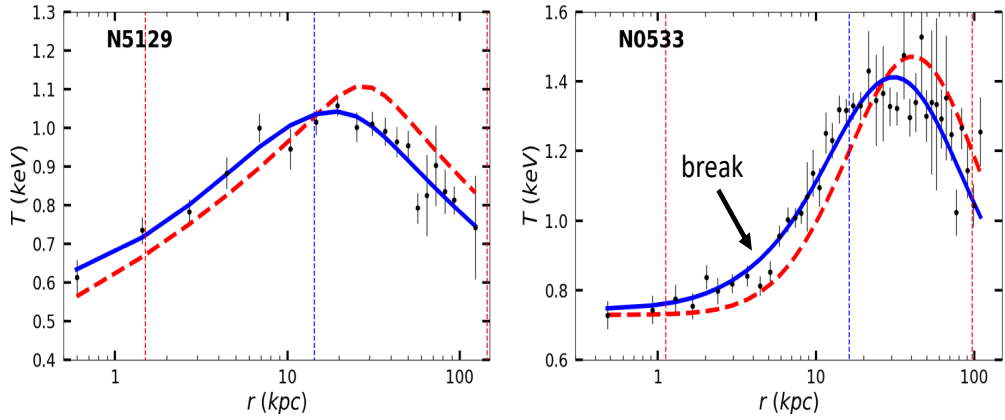

Figure 2. Hybrid-bump temperature profiles, (left) one that shows a peak with one break and (right) another that shows a 2$^{nd}$ break in the slope at small radii, r < $R_{MAX}$. All symbols and colors are the same as in Figure 1.

## 3.2 Hybrid-Dip

This is the 2$^{nd}$ largest with 13 galaxies or 22% in our sample. Opposite to the hybrid-bump type, the profiles in this hybrid-dip type have a negative temperature gradient at r < $R_{MIN}$ and a positive gradient at r > $R_{MIN}$ with a single temperature minimum at r = $R_{MIN}$ in the middle of the observable radius range. In this type, the dip is found at $T_{MIN}$ = 0.49 ± 0.16 keV and $R_{MIN}$ = 3.4 ± 2.1 kpc, which is considerably cooler and smaller than $T_{MAX}$ and $R_{MAX}$ of the hybrid-bump type (see Table 2 and section 4 for more discussions). NGC 499, which is likely a small group, has the largest $R_{MIN}$ (9.9 kpc) in this type.

In terms of the temperature gradient at r > $R_{MIN}$, there may be two cases. Some galaxies (NGC 1332, NGC 4278) have a slope (~0.3; see section 4), which is similar to that of the hybrid-bump type, while others (e.g., NGC 499, NGC 1404) have a considerably steeper slope. The latter case is a non-BCG galaxy (or a subgroup) embedded in the hotter gas (see Appendix A for individual galaxies and section 5 for more discussions).

### 3.3 Double-Break

There are four galaxies with double-breaks in their temperature profile, i.e., a combination of hybrid-bump and hybrid-dip profile types. They show a dip at small radii and a peak at larger radii. The temperature gradient is negative, positive, then negative from the center to the outskirts. There is no case with an opposite combination, i.e., a peak in small radii and a dip in large radii.

### 3.4 Positive

There are six galaxies with a positive gradient with no change in the gradient sign within the observable radius range. The temperature increases monotonically from the center to the outskirts. Most galaxies in the type may be similar to the hybrid-bump type, but we may not observe the outer temperature drop, because they are embedded inside the hotter groups/clusters (e.g., NGC 4472 in the Virgo cluster) and/or the outer region is not observed due to the limited ACIS field-of-view (see more discussion in section 5.)

### 3.5 Negative

There are eight galaxies with a negative gradient with no change in the gradient sign within the observable radius range. Opposite to the positive case, the temperature decreases monotonically from the center to the outskirts. Some galaxies (e.g., NGC 4382) in this type may be similar to the hybrid-dip type, but we cannot measure the outer region where the temperature slope changes the sign to be positive, because the X-ray emission is too faint (see more discussion in section 5.) However, NGC 6482 is the most obvious example of this negative type with a continuously decreasing T in a wide radius range (from a few kpc to ~ 60 kpc)

### 3.6 Irregular

There are three cases where the T profile does not fit any type listed above. A good example is NGC 3402 (= NGC 3411), where cooler gas forms a shell-like structure at 20-40 kpc, which is surrounded by inner and outer hotter gas with a relatively constant temperature (~1 keV) (O'Sullivan et al. 2007). Some profiles in the hybrid-dip type may look similar to these cases with a cooler ring, but in general, the slope change in the hybrid-dip type is smoother in a wider radius range than those of NGC 3402. Because of the irregular nature of the profile, the best-fit line in Appendix A does not represent reality.

    The second case is NGC 5813, where the hot gas morphology exhibits three sets of nested co-aligned cavities and shocks (Randall et al. 2011 and 2015). Consequently, the temperature profile shows multiple peaks (at ~1 kpc, ~10 kpc) and dips (at ~3 kpc and ~20 kpc). Another peak at r ~ 50 kpc (similar to the peak in hybrid-bump) is identified by XMM-Newton data (in prep.)

Again, the best-fit line in Appendix A does not represent reality.

The last case of the irregular type is NGC 7618, which is a well-known sloshing system (Kraft et al. 2006, Roediger et al. 2012). The pronounced spiral-like features, likely caused by UGC 12491, redistribute the cooler gas resulting in the complex temperature profile with a positive gradient at r < ~2 kpc, a negative gradient at r = 2- 30 kpc, then a positive gradient at r < ~30 kpc. The shape may look like a reversed double-break type, but the temperature is likely to drop again at the outskirts, making it different from the typical double-break type.

In summary, we identified 26 hybrid-bump, 13 hybrid-dip, 4 double-break, 6 positive, 8 negative types. The remaining 3 are irregular for their specific reason.

## 4. CHARACTERISTICS OF TEMPERATURE PROFILES

We find that ETGs in our sample show complex thermal structures. None of them can be described as isothermal. While some (23%) have a single temperature gradient (positive and negative types), the majority of ETGs have multiple gradients with one or two breaks (72%), having both positive and negative gradients in their radial profiles (hybrid-bump, hybrid-dip, and double-break types).

Table 2. Means and standard deviations of the peaks and dips in individual profile types

| | N | $<T_{MIN}>$ σ | $<T_{MAX}>$ σ | $<R_{MIN}>$ σ | $<R_{MAX}>$ σ |
|---|---|---|---|---|---|
| Hybrid-Bump | 26 | | 1.4 0.33 | | 34.8 19.7 |
| Hybrid-Dip | 13 | 0.47 0.16 | | 3.4 2.1 | |
| Double-Break | 4 | 0.75 0.12 | 1.3 0.13 | 1.4 0.7 | 32.0 9.1 |

| | N | $<T_{BREAK}>$ σ | $<R_{BREAK}>$ σ |
|---|---|---|---|
| Hybrid-Bump | 17 | 0.86 0.16 | 4.9 2.4 |

$R_{MAX}$ and $T_{MAX}$ are the galacto-centric distance and temperature at the peak of the T bump.
$R_{MIN}$ and $T_{MIN}$ are the galacto-centric distance and temperature at the bottom of the T dip.
$R_{BREAK}$ and $T_{BREAK}$ are the galacto-centric distance and temperature at the inner break of the 17 hybrid-bump type galaxies.

### 4.1 The Temperature Peaks and Dips in the R-T Plane

To examine the observed characteristics of the temperature profiles in our sample quantitatively, we compare the bumps and dips of the profiles in terms of their temperature and galacto-centric distance. In the left panel of Figure 3, we plot the temperatures of the peaks (red upward triangles) of the hybrid-bump and the dips (blue downward triangles) of the hybrid-dip types against their radii. In this R-T plane, the peaks and dips are clustered in two distinct locations: the dips are found at the lower-left corner and the peaks at the upper-right corner. In other words, the peaks are found at higher T and larger R than the dips. Quantitatively, the dip of the hybrid-dip type has a $T_{MIN}$ (~0.5 keV), which is always lower than the $T_{MAX}$ (~1.4 keV) of the peak in the hybrid-bump type with no exception. The mean and standard deviation of the dips and peaks are listed in Table 2 and marked in Figure 3 by blue and red crosses, respectively. The difference between the two mean temperatures is significant at the 2.5σ confidence level. The galacto-centric distance of the dip ($R_{MIN}$ ~ 3.4 kpc) is smaller than that of the peak ($R_{MAX}$ ~ 35 kpc) with a small number of exceptions. In one case of the hybrid-dip type, the dip is at ~10 kpc ($R_{MIN}$ = 9.9 kpc for NGC 499), and in one case of the hybrid-bump type, the peak is inside 10 kpc ($R_{MAX}$ = 7 kpc for NGC 6861). The difference between the two mean radii is significant at the 1.6σ level. Based on the 2-dimensional Kolmogorov–Smirnov test (Fasano & Franceschini 1987), the null hypothesis probability that two subsamples are originated from the same parent population is 4.3 x $10^{-8}$, suggesting that the dip and peak are two distinct characteristics.

There are a small number (4) of interesting double-break cases where the temperature profile has both a peak and a dip. We mark them in the left panel of Figure 3 by the green upward and downward triangles, respectively. In all 4 cases we find $R_{MAX} > R_{MIN}$ and $T_{MAX} > T_{MIN}$. The peak and dip of this type are similar in their radius and temperature to the peak of the hybrid-bump type and the dip of the hybrid-dip types, respectively, suggesting that they, albeit rare, can provide important clues on possible connections between the hybrid-dip and hybrid-bump types (see section 5).

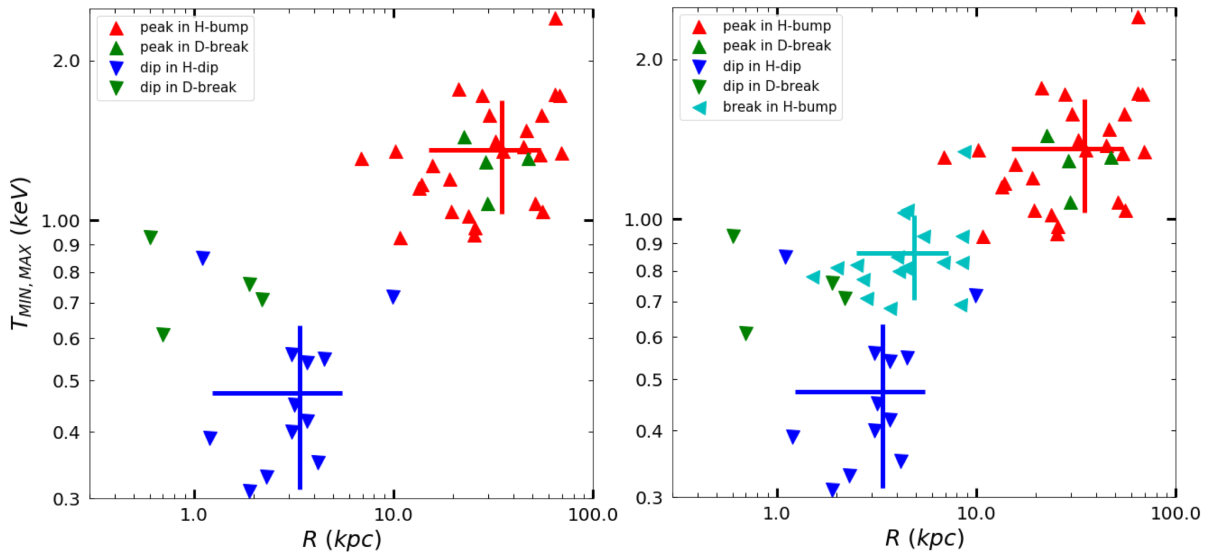

Figure 3. Comparison of the peaks and dips of the temperature profiles in term of their temperature and galacto-centric distance. The crossbars indicate the group mean and standard deviation of individual profile types, marked by the same color.

As described in section 3.1 and Figure 2, 17 of 26 galaxies of the hybrid-bump type have an inner temperature break with a slope being flatter at small radii (r < $R_{BREAK}$). These 17 breaks of the hybrid-bump type are plotted in the right panel of Figure 3 (left-pointing cyan triangles). The mean and standard deviation of these breaks are marked by a cyan cross (see also Table 2). Their galacto-centric distances are similar to those of the dips of the hybrid-dip type. Their temperatures are slightly higher than those of the dips of the hybrid-dip type.

The similar location of the inner break of the hybrid-bump type and the dip of the hybrid-dip and double-break types suggests that they may be linked, possibly by a similar origin. To further investigate this possibility, we plot the dips, peaks, and breaks in a rescaled R-T plane. In Figure 4, $T_{MIN}$, $T_{BREAK}$, and $T_{MAX}$ are rescaled by $T_{GAS}$ for each galaxy, which is the gas temperature determined by a single spectral fitting (with the spectra extracted from the entire region). $T_{GAS}$ is taken from the Chandra Galaxy Atlas (see Kim et al. 2019a and references therein). The galacto-centric distance is also scaled by the viral radius ($R_{VIR}$) which is determined by the scaling relation given by Helsdon and Ponman (2003),

$$R_{VIR} = 0.81 \, (T_{GAS} / 1 \text{ keV})^{1/2} \text{ Mpc.}$$

We list $R_{VIR}$ of individual galaxies in Table 1. Remarkably, the inner breaks and the dips are statistically identical in this scaled R-T plane. This strongly suggests that the hybrid-bump, double-break, and hybrid-dip types are related. Furthermore, given that the lack of a temperature peak in the hybrid-dip type may be caused by the observational limitation and/or selection effects, these three types may share the common temperature profile shape (see section 5). We also find that the scatters in R and T of the peak, and the dip/break are comparable, typically 0.1 – 0.2 dex.

The thin lines in Figure 4 connect the inner break and the peak for the 17 hybrid-bump type galaxies and the dip and the peak for the four double-break type galaxies. The slopes are similar. The mean slope is 0.3 ± 0.1 for these 21 galaxies, again suggesting common characteristics in their temperature profile. As noted in section 3.2, the slope of hybrid-dip type galaxies at r > $R_{MIN}$ varies widely, because some galaxies are embedded inside the hotter gas of the groups and clusters.

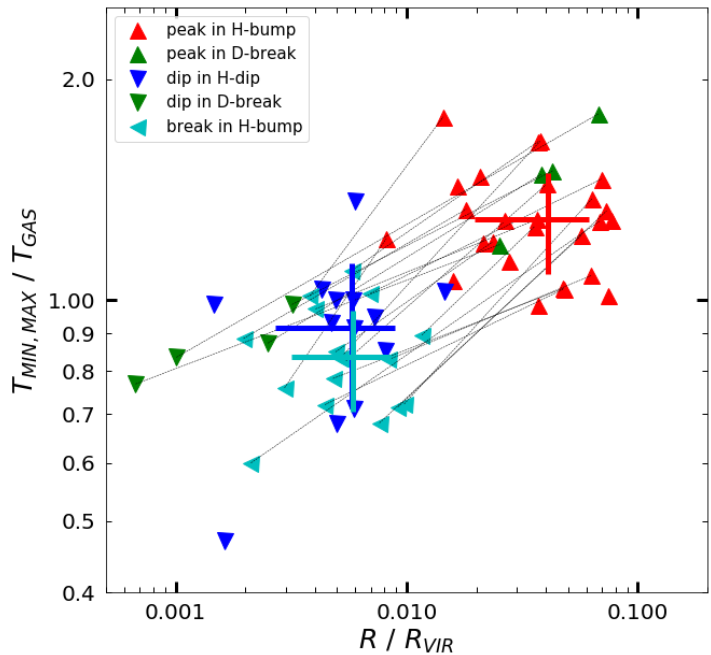

Figure 4. Same as Figure 3, but R and T are rescaled in a log scale by $R_{VIR}$ and $T_{GAS}$. The thin lines connect the peak and the inner break (or the dip) for 21 individual galaxies. The mean slope is 0.3 ± 0.1.

In summary, we find an obvious break of the temperature gradient in the three-quarters of our sample. Of these, we identify a peak at the outer region in 30 ETGs, and a dip (or an inner break) at the inner region in 35 ETGs. We find two breaks in 21 ETGs. The temperature peaks are found at r ~ 35 kpc (or ~0.04 $R_{VIR}$) on average, and the temperature dips/breaks are at r ~ 4 kpc (or ~ 0.006 $R_{VIR}$) on average. The slope between the peak and the dip (or the inner break) is ~0.3.

## 4.2 Different Types in the $L_{X,GAS}$-$T_{GAS}$ Plane

To understand the characteristics of ETGs in individual temperature profile types better, we look at the global properties of the hot gas. Figure 5 shows their $L_{X,GAS}$ -$T_{GAS}$ relations. $L_{X,GAS}$ is the X-ray luminosity in 0.3-8 keV from the entire hot ISM and $T_{GAS}$ is the gas temperature determined by the spectra extracted from the entire region or similar to the luminosity-weighted mean temperature in a logarithmic scale. We take $L_{X,GAS}$ and $T_{GAS}$ from the Chandra Galaxy Atlas (see Kim et al. 2019a and references therein). $L_{X,GAS}$ was measured primarily with Chandra data. For galaxies with extended halos, ROSAT or XMM-Newton results were taken from the literature. While the entire sample shows a tight positive correlation as previously known (e.g., Boroson et al. 2011; Kim & Fabbiano 2015; Goulding et al. 2016; Babyk et al. 2018), the locations of different profile types in this L-T plane convey useful information. On the left panel, we compare the hybrid-bump and hybrid-dip types. The hybrid-bump type galaxies (red upward triangles) are preferentially found in the upper right corner and the hybrid-dip type galaxies (blue downward triangles) in the lower-left corner, i.e., the hybrid-bump type galaxies host hotter and more luminous gas than the hybrid-dip type galaxies. The means and standard deviations of the subsamples are marked by crossbars with the same colors as the data points in Fig 5 (see also Table 3.) Based on the 2-dimensional Kolmogorov–Smirnov test (Fasano & Franceschini 1987), the probability that two subsamples are originated from the same parent population is 2.7 x $10^{-5}$, indicating they are significantly different at the 4σ level.

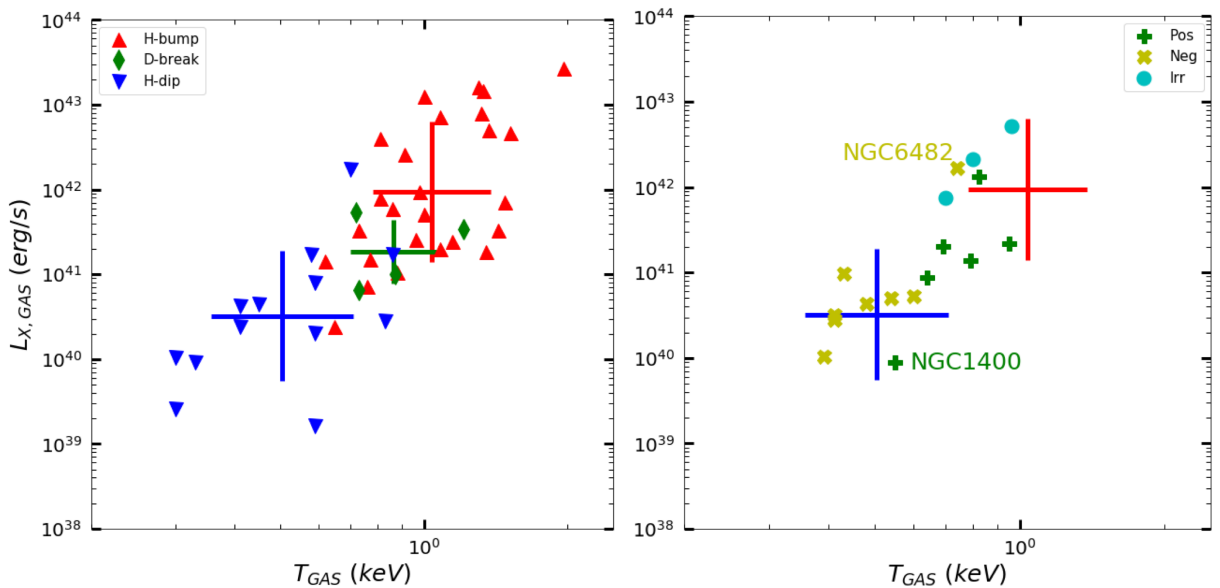

Figure 5. Comparison of the different temperature profile types in terms of their gas temperature ($T_{GAS}$) and luminosity ($L_{X,GAS}$). (a) The hybrid-bump type (red upward triangles) and hybrid-dip type (blue downward triangles) are seen in two separate locations. The double-break type (green diamonds) falls in the middle. The crossbars with the same colors indicate the mean and standard deviation of each type. (b) The other three types are compared. The red and blue crossbars are the same as in (a).

Also plotted in the left panel of Figure 5 is the double-break type (green diamonds). Its sample mean and deviation is marked by the green cross. They lie at the lower end of the hybrid-bump type.

On the right panel of Figure 5, we show the other three types: positive, negative and irregular. For comparison, the two crossbars from the left panel are over-plotted here. The negative type is found in the lower-left corner which is similar to the hybrid-dip. One exception is NGC 6482 which is comparable to the hybrid-bump. On the other hand, the positive type galaxies host hotter and more luminous gas than the negative type galaxies (and hybrid-dip), but they are not as hot/luminous as the typical hybrid-bump type. One exception is NGC 1400 which is comparable with the hybrid-dip.

```
Table 3. Means and standard deviations of Lx,GAS and TGAS
-------------------------------------------------------
              N    <log(Lx,GAS)>  σ        <TGAS>   σ
-------------------------------------------------------
Hybrid-Bump   26     41.97       0.83       1.04    0.34
Hybrid-Dip    13     40.61       0.82       0.52    0.21
Double-Break  4      41.27       0.37       0.86    0.20

Positive      6      41.15       0.68       0.74    0.14
Negative      8      40.78       0.60       0.49    0.11
Irregular     3      42.30       0.40       0.82    0.13
-------------------------------------------------------
```

In summary, the hottest and most luminous galaxies tend to be hybrid-bump, i.e., galaxies with a peak in their temperature profile. On the other hand, the coolest and least luminous galaxies tend to hybrid-dip or negative types, i.e., galaxies with a negative T gradient in the inner region. The double-break and positive types seem to bridge the gap between the hybrid-bump and hybrid-dip types, being at intermediate temperature and luminosity. We note that given the selection effect of our sample, the hybrid-dip and negative types are likely under-represented because they are X-ray faint galaxies (see section 5).

## 5. IS THERE A UNIVERSAL TEMPERATURE PROFILE?

As described in section 3, the most common type is hybrid-bump. Together with the double-break type, they comprise 50% of our sample (30 out of 60 galaxies). The main characteristic feature in their temperature profiles is that the temperature peaks at $R_{MAX}$ and decreases both inward and outward from the peak, i.e., the temperature gradient is positive between $R_{MIN}$ (or $R_{BREAK}$) and $R_{MAX}$ and negative in the outskirts (r > $R_{MAX}$). At smaller radii (at r < $R_{MIN}$ or r < $R_{BREAK}$), the temperature gradient (1) remains constant (positive) in 9 out of 26 hybrid-bump type galaxies, (2) breaks at $R_{BREAK}$ (becoming flatter, close to 0) in 17 of 26 hybrid-bump type galaxies, or (3) changes its sign to negative in 4 double-break type galaxies.

Motivated by these characteristic temperature profiles (and the reasons described below for other types), we further explore the possibility of a *universal* temperature profile of the hot ISM in ETGs. To test whether the T profiles can be adequately scaled to identify the universal profile, in Figure 6 (left panel), we plot the temperature of all galaxies in the hybrid-bump and double-break types. The temperature is scaled by $T_{MAX}$ (in Figure 3) and the radius is scaled by the viral radius ($R_{VIR}$). Once scaled properly, the temperature profile follows a common shape qualitatively as well as quantitatively in the full radius range except in the inner region (r < $R_{MIN}$). The temperature peak at ~0.04 $R_{VIR}$, the positive gradient with a slope of ~0.3 (± 0.1) at $R_{MIN}$ < r < $R_{MAX}$, and the negative gradient with a relatively steeper slope of ~0.5 (loosely defined, though) at r > $R_{MAX}$ are identifiable. The scatter is large in the inner region due to a wide range of the core T gradient (see section 6). In the right panel of Figure 6, we show a schematic diagram of this temperature profile shape, indicating the bump, the dip or the inner break and the slope of 0.3 between $R_{MIN}$ (or $R_{BREAK}$) and $R_{MAX}$. The outer slope (at r > $R_{MAX}$) will be measured more reliably in the next paper with additional XMM-Newton archival data.

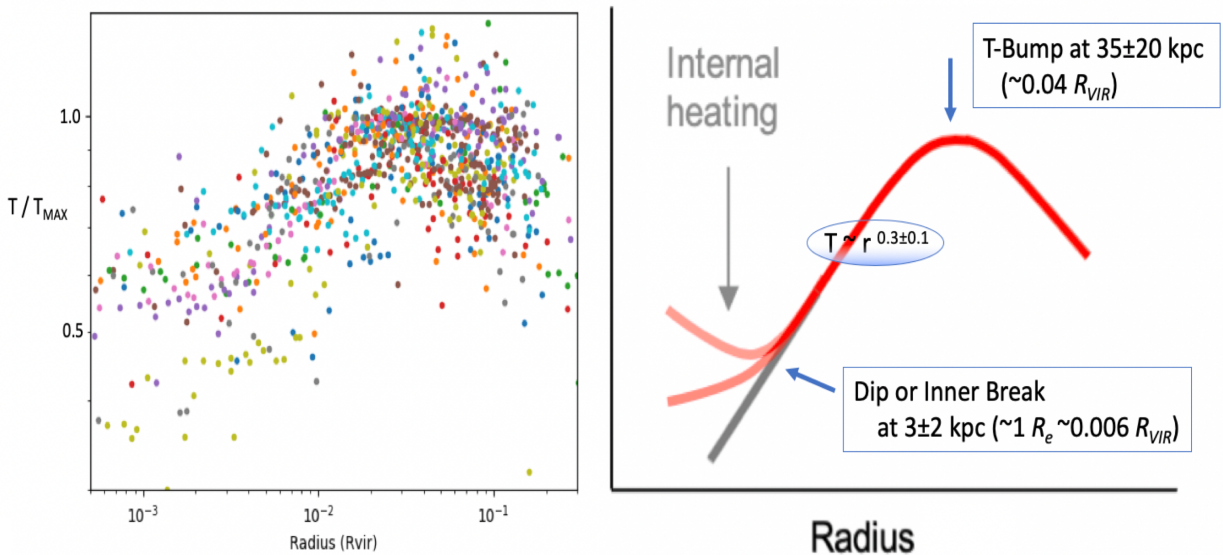

Figure 6. (left) Temperature (scaled by $T_{MAX}$) against radius (scaled by $R_{VIR}$) for all galaxies in the hybrid-bump and double-break types. (right) A schematic diagram of the proposed "universal" temperature profile.

The second most popular type is hybrid-dip. In section 4.1 and Figure 4, we demonstrate that when properly scaled, $T_{MIN}$ and $R_{MIN}$ of this type are close to $T_{BREAK}$ and $R_{BREAK}$ of the hybrid-bump type and $T_{MIN}$ and $R_{MIN}$ of the double-break type, suggesting that this type may follow the universal profile (that of the double-break type) in Figure 6. The only difference is that this hybrid-dip type does not have a temperature peak and a drop after the peak, but continuously increases to the observable radius limit. The lack of the peak may be caused by observational limitations and selection effects such that we may not see the ISM temperature declining at large radii. We consider the following three possible reasons. First, the hybrid-dip type consists primarily of hot gas poor ETGs (see Figure 5). Because of their small amount of hot gas, their X-ray emission is not detected at large radii. This type might be similar to the double-break if the hot gas could be detected in the outer region. NGC 1332, NGC 3923, and NGC 4278 are good examples as their hot gas is detected only out to r ~ 20 kpc. Second, some of the hybrid-dip type galaxies are inside groups or clusters, i.e., they are satellite galaxies embedded in hotter IGM/ICM. In this case, the X-ray emission from the outer region of this satellite galaxy is dominated by the hotter gas such that the temperature at large radii remains high. Again, this system might be similar to the double-break type if the hot ISM could be separated from the hotter IGM/ICM. NGC 1380, NGC 1387, NGC 1404 (inside the Fornax cluster centered on NGC 1399), NGC 499 (a subgroup possibly merging with the NGC 507 group), NGC 4552 and NGC 4649 (inside the Virgo cluster) are such examples. One of the distinct characteristics of these galaxies is the steeper positive T gradients at r > $R_{MIN}$ than the other galaxies in the hybrid-dip type (see Appendix A). Third, the detector field-of-view (fov) is not large enough and the hot gas properties in the outer region are not well constrained. NGC 4342 is an example. The maximum radius where the hot gas emission is reliably detected with an azimuthal coverage larger than 95% is 20 kpc (Kim et al. 2019a). See the vertical line in Appendix A which indicates this limiting radius. Adding the hybrid-dip type, we can apply the universal temperature profile to 43 out of 60 ETGs or 72% of our sample.

To illustrate the different types that may follow the proposed universal temperature profile in Figure 6, we compare the fiducial temperature profile and observed deviations from this fiducial profile in Figure 7. The top row shows the hybrid-bump type with no inner break (the temperature gradient is constant at r < $R_{MAX}$). The 2nd row shows the hybrid-bump type, but with an inner break (the temperature gradient is flatter at r < $R_{BREAK}$), which indicates some internal heating as illustrated in the right-hand panel. The 3rd row shows the double-break type where the temperature gradient is negative at r < $R_{MIN}$, which indicates even stronger internal heating. The 4th row shows the hybrid-dip type where the profile ends before it reaches the peak. This type may be similar to the double-break type (the 3rd row) and the lack of the peak may be caused by observational limitations and selection effects as described in the above.

Further extending the idea of the universal profile, we may explain the positive type (the bottom row in Figure 7) as those of the hybrid-bump type (the top two rows in Figure 7), but embedded in the hotter IGM/ICM such that the temperature peak of their own systems cannot be seen. This system might be similar to the hybrid-bump if the hot ISM could be separated from the hotter IGM/ICM. NGC 1400 (in the NGC 1407 group), NGC 4472, NGC 4406 (both in the Virgo cluster) and NGC 7626 (in the NGC 7619 group) are such an example. In particular, the temperature profile of NGC 4472 (the bottom panel of Figure 7) shows that the profile starts off relatively flat before rising sharply at r ~ 4 kpc. At r ~ 12kpc, the profile plateaus before rising sharply again due to the hot intra-cluster medium (ICM). The inner break at ~4 kpc may be the same as the inner break in hybrid-bump type and the plateau at 12 kpc could be due to the ISM being washed out by the hotter ICM of the Virgo cluster. If the positive type is similar to the hybrid-bump type but embedded inside a hotter environment, the universal temperature profile can be applied to 49 out of 60 ETGs or 82% of our sample.

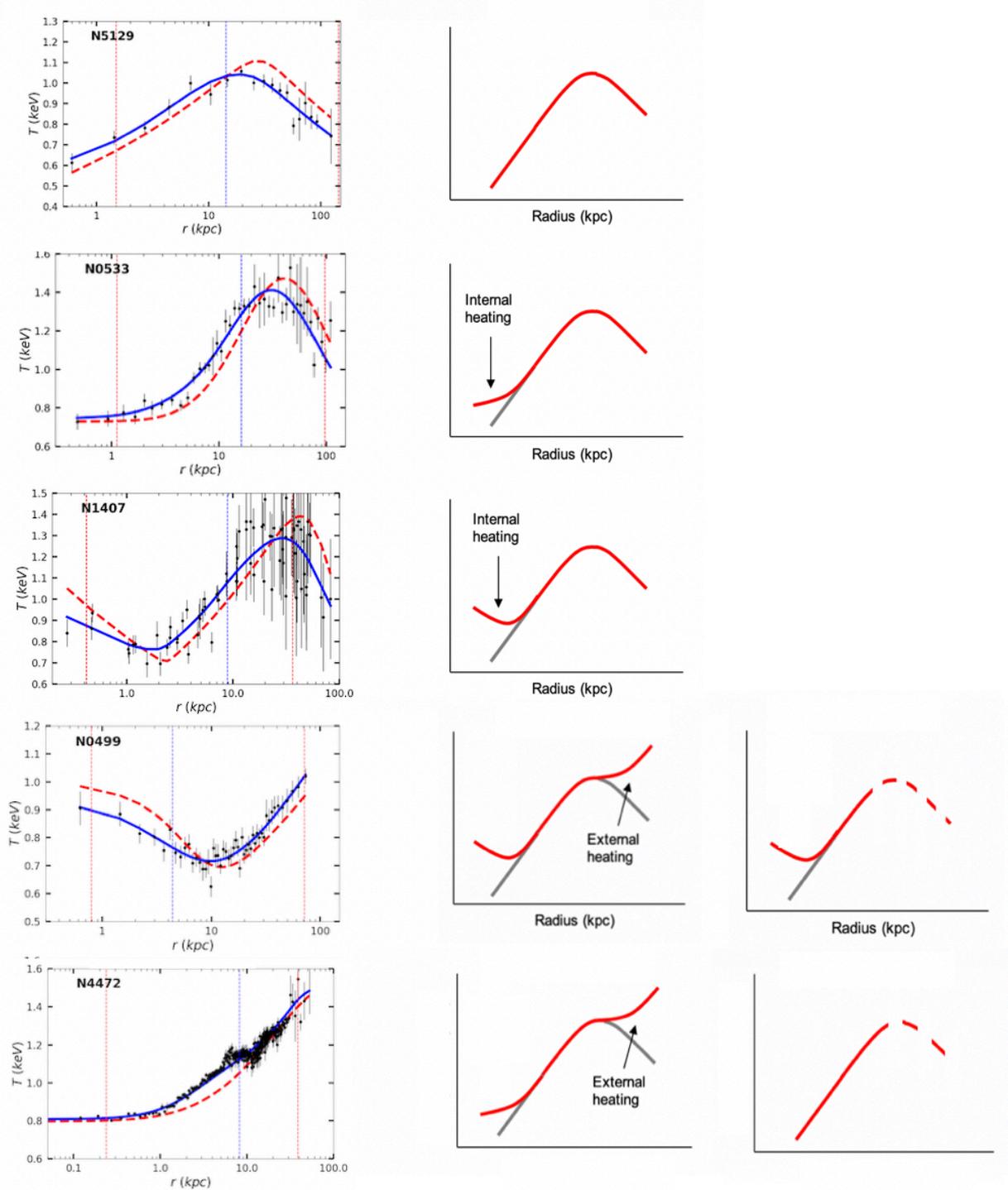

Figure 7. The left-hand panels show examples of hybrid-bump (without a break), hybrid-bump (with an inner break), double-break, hybrid-dip, and positive temperature profiles (top to bottom). The right-hand panels show a graphic illustration of how each profile may be explained by a "universal" profile. The grey curve represents the fiducial profile and the red curve shows the observed profile with deviation from the fiducial profile.

The most obvious exception to the proposed universal temperature profile is the negative types which do not fit the above explanations. Since this type is generally a small system with low $T_{GAS}$ and $L_{X,GAS}$ (see Figure 5), one might consider this is an extreme case of a hybrid-dip type being truncated at $r < R_{MIN}$ such that we could only see the inner negative gradient without the temperature minimum. NGC 1316 and NGC 4382 (see Appendix A) could be such an example as they show a hint of temperature rise at large radii, albeit with large uncertainty. However, most galaxies in this type have the extended hot gas detected out to at least $r \sim 20$ kpc. The extent of the hot gas is considerably larger than $<R_{MIN}> \sim 4$ kpc, even larger than the largest $R_{MIN}$ (~10 kpc in NGC 499). The most obvious examples are NGC 4125 and NGC 6482 where the temperature is constrained out to $r = 40\text{-}50$ kpc with no sign of a temperature gradient reversal. Also note that NGC 6482 has $L_{X,GAS}$ and $T_{GAS}$ which are comparable to that of the hybrid-bump type. The negative type may be similar to those of the non-cool core (NCC) clusters, which are often found in disturbed systems, like mergers and could have an irregular or declining temperature profile. However, NGC 6482 is one of the known relaxed, fossil systems (Khosroshahi et al. 2004), so it is quite different from the NCC clusters.

In summary, the temperature profiles of at least 72% of our ETG sample (hybrid-bump, hybrid-dip, and double-break types) can be described by a universal temperature profile. In addition, if we consider that the positive types are embedded in a hotter IGM/ICM such that the temperature peak of the galaxy is not detected, then 82% of our sample fit the universal temperature profile. The negative (13% of our sample) and irregular (5%) types do not fit the universal profile and require another explanation.

## 6. DISCUSSION

Analyzing the temperature profile of 60 ETGs, we have identified six profile types: in order of decreasing frequency of each type, hybrid-bump, hybrid-dip, negative, positive, double-break, and irregular. The hybrid-bump type is the majority (43%) in our sample, but the hybrid-dip and negative types are likely under-represented. Considering selection effects and observational limits, we find common characteristics of the temperature profile among 72% (possibly up to 82%), a temperature peaking at $R_{MAX}$ and declining inward and outward, except the inner temperature gradient inside $R_{MIN}$ (or $R_{BREAK}$) which can vary widely. We note that because ETGs and small groups are not easy to separate, our sample includes some small groups (but only those with $T_{GAS} < 1.5$ keV).

X-ray studies of the hot gas in groups and clusters often separate them into two groups, cool cores (CC) and non-cool cores (NCC) (e.g., Molendi & Pizzolato 2001; Sanderson et al. 2009; Hudson et al. 2010). CCs are generally relaxed systems with cuspy cores and typically have higher metallicity and lower central temperature (T peaking at ~ 0.1 $R_{VIR}$ and declining toward the center) and lower entropy than NCCs. NCCs are often disturbed systems, suggesting that they originate from mergers in denser environments or infalling substructures (McCarthy et al. 2011; Gaspari et al. 2014). Most ETGs in our sample belong to CC. As described in section 3, those galaxies in the hybrid-bump, double-break, hybrid-dip, and positive have a positive temperature gradient (or a declining temperature profile toward the center) in the region between $R_{MIN}$ (or $R_{BREAK}$) and $R_{MAX}$ (or the observation limit). In this respect, 82% of our sample belong to CC, regardless of their inner temperature gradients (positive or negative). When the inner temperature gradient breaks, that occurs inside the cool core, i.e., $R_{MIN}$ is always smaller than $R_{MAX}$ ($R_{MIN} \sim 0.15 \times R_{MAX}$).

The negative type galaxies in our sample may be an analog of the NCC clusters. NGC 1316 is a good example because it exhibits many signs of recent mergers (e.g., Schweizer 1980; Kim & Fabbiano 2003). However, the most obvious case with the temperature declining in a wide range of radius, NGC 6842, is known to be relaxed with no nearby possible perturber (a fossil system), in contrast to the disturbed NCC clusters.

T increases toward the center for 42% of our sample. We refer to these as *hot core* (HC) hereafter. Note that HC is not the opposite of CC. HC, if it exists, stays inside CC. Unlike the CC clusters where the hot gas properties are primarily controlled by the gravity, non-gravitational baryonic physics plays an important role in ETGs, most significantly in the inner region, possibly by additional heating (e.g., stellar feedback, AGN feedback). The stronger non-gravitational effect was previously demonstrated in the $L_{X,GAS}$ - $T_{GAS}$ relation which is considerably steeper in ETGs ($L \sim T^{4.5}$) than in clusters ($L \sim T^3$) (Kim & Fabbiano 2015). The same effect may be reflected in our finding that the peak of the temperature bump is found at a considerably smaller R (~0.04 $R_{VIR}$) that that (~0.1 $R_{VIR}$) in groups and clusters and that the mean slope between the bumps and the inner break (or the dip) is ~0.3 which is small than that (0.5) expected from the pure cooling flow (Gaspari et al. 2012) and that (~0.4) of CC clusters (e.g., Sanderson et al. 2006).

We note that the distribution of the inner T gradient is not bimodal, i.e., the slope varies continuously in our sample. The presence or absence of HC does not separate ETGs into two distinct groups, in contrast to the bimodality between CC and NCC clusters (e.g., Sanderson et al. 2009).

HC is often accompanied by a flattened surface brightness profile at the core which in turn causes a flatter density profile (less steeply increasing toward the center, forming a core) and a flatter entropy profile (less steeply declining toward the center forming an entropy floor) than those without HC. Note that the density profile depends on accurate abundance measurements such that the density (and entropy) would look steeper (flatter) if a radial abundance gradient is ignored. We will present the full analysis of density and entropy profiles in the next paper.

## 6.1 The definition of $\nabla T_{CORE}$

In sections 4 and 5, we show that the inner temperature gradient varies widely from one galaxy to another. To further investigate the inner temperature profile inside $R_{MIN}$ (or $R_{BREAK}$) which is 3 (5) kpc on average, we measure the core temperature gradient ($\nabla T_{CORE}$ = d logT /d logR) at r = 0.5 – 1 kpc to explore the innermost hot gas property, but set a constraint on the minimum radius of 3" to confidently remove any potential contamination by the nuclear emission. Out of 52 galaxies for which we measure $\nabla T_{CORE}$, 25 galaxies have a negative $\nabla T_{CORE}$ with HC (hybrid-dip, double-break, and negative types) and 27 galaxies have a positive $\nabla T_{CORE}$ with no HC (hybrid-bump and positive types).

The presence of HC was previously known in a small number of ETGs (e.g., NGC 4278, Pellegrini et al. 2012; NGC 4649, Humphrey et al. 2006 and Paggi et al. 2014; NGC 4552, Machacek et al. 2006), and in some groups (NGC 777 and NGC 5982, O'Sullivan et al. 2017). In this paper, we have systematically searched for HC and investigate them for the first time in a statistical manner. In particular, we explore the stellar feedback, AGN feedback and gravitational heating for the cause of the HC in the following sections. DS08 have considered the outer temperature gradient associated with the environmental effect. We will address this question in the next paper with additional XMM-Newton archival data.

## 6.2 $\nabla T_{CORE}$ - Global Hot Gas Properties

To investigate how the inner temperature gradient (or the presence or absence of HC) is related to other galaxy properties, we first plot $\nabla T_{CORE}$ against $L_{X,GAS}$ and $T_{GAS}$ in Figure 8. The different T-profile types are marked by different symbols and colors as in Figure 5. As expected from Figure 5, galaxies with a positive $\nabla T_{CORE}$ (i.e., hybrid-bump and positive types) have higher $T_{GAS}$ and $L_{X,GAS}$ than those with a negative $\nabla T_{CORE}$ (i.e., hybrid-dip and negative types). We applied the linear regression method[6] given by Kelly (2007). We also applied the Pearson and Spearman correlation tests taken from the scipy statistics package[7] to estimate the p-value for the null hypothesis. Based on the linear regression, we find a best-fit slope of 0.68 ± 0.1 and 0.13 ± 0.02 for $T_{GAS}$ and $L_{GAS}$, respectively. Based on the two correlation tests, $\nabla T_{CORE}$ is correlated with $T_{GAS}$ and $L_{X,GAS}$ at the 5-6 $\sigma$ level with the p-value ranging from $10^{-8}$ to $10^{-10}$. The results are summarized in Table 4.

Given that the overall trends are driven by the fact that those with HC (negative gradient) are cooler and less luminous than those without HC, we further test the correlations for those with HC (HC sub-group) and without HC (nHC sub-group), separately. The correlation becomes less tight in both subsamples, but is still significant (p-value = 0.06 – 0.0007) for the HC sub-group, while there is almost no correlation (p-value = 0.1-0.2) for the nHC sub-group. The correlation tests suggest that the presence of HC is more pronounced in smaller galaxies with lower $T_{GAS}$ and $L_{X,GAS}$. Regarding the inner break, its presence ($\nabla T_{CORE} \sim 0$) or absence ($\nabla T_{CORE} > 0$) among the hybrid-bump type has no preference in $T_{GAS}$ and $L_{X,GAS}$, once they are large enough.

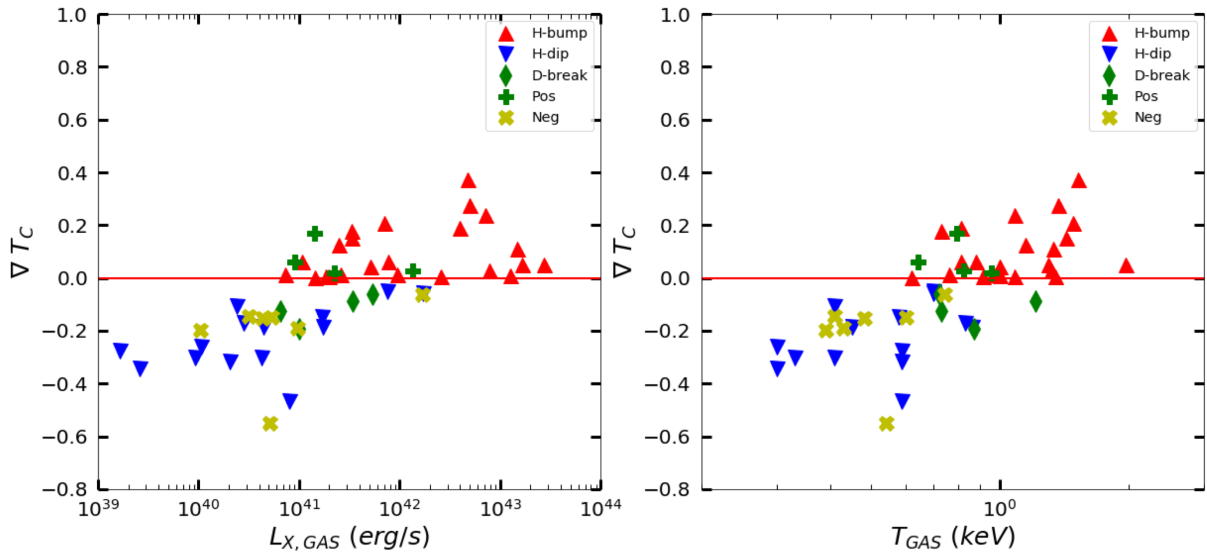

Figure 8. The inner T-gradient ($\nabla T_{CORE}$) against the temperature ($T_{GAS}$) and X-ray luminosity ($L_{X,GAS}$) of the hot gas. All symbols are the same as in Figure 5.

---

[6] https://github.com/jmeyers314/linmix
[7] http://www.scipy.org

We further explore the relation of $\nabla T_{CORE}$ with other important galaxy parameters that may affect the core temperature gradient, including (section 6.3) $L_K$ indicating the stellar mass, the mean age indicating recent star formation; (section 6.4) the radio luminosity at 1.4 GHz and the hard X-ray core luminosity both indicating the strength of the nuclear activity; (section 6.5) the central stellar velocity dispersion indicating the dynamical mass in the core, and the total mass indicating the virial mass. All quantities and their sources are listed in Table 1.

### 6.3 $\nabla T_{CORE}$ - Stellar Properties

The left panel of Figure 9 shows the relation between the core temperature gradient and the stellar K-band luminosity ($L_K$). $L_K$ is a good proxy for stellar mass because the stellar mass to light ratio for ETGs ($M_{STAR} / L_K$ in unit of $M_\odot/L_\odot$) is close to 1 (e.g., Bell et al. 2003). $\nabla T_{CORE}$ is closely correlated to the stellar K-band luminosity at the 5$\sigma$ level. As in Figure 8, the fact that the hybrid-bump and positive types are generally larger than the hybrid-dip and negative types drives the overall trend. For the HC subgroup, the correlation remains strong with the p-value of 0.01-0.02 or at the ~2.5$\sigma$ level, but for the nHC subgroup, the high p-value (0.1-0.2) indicates almost no correlation. This further suggests that the heating source in the core is most efficient in the smallest system.

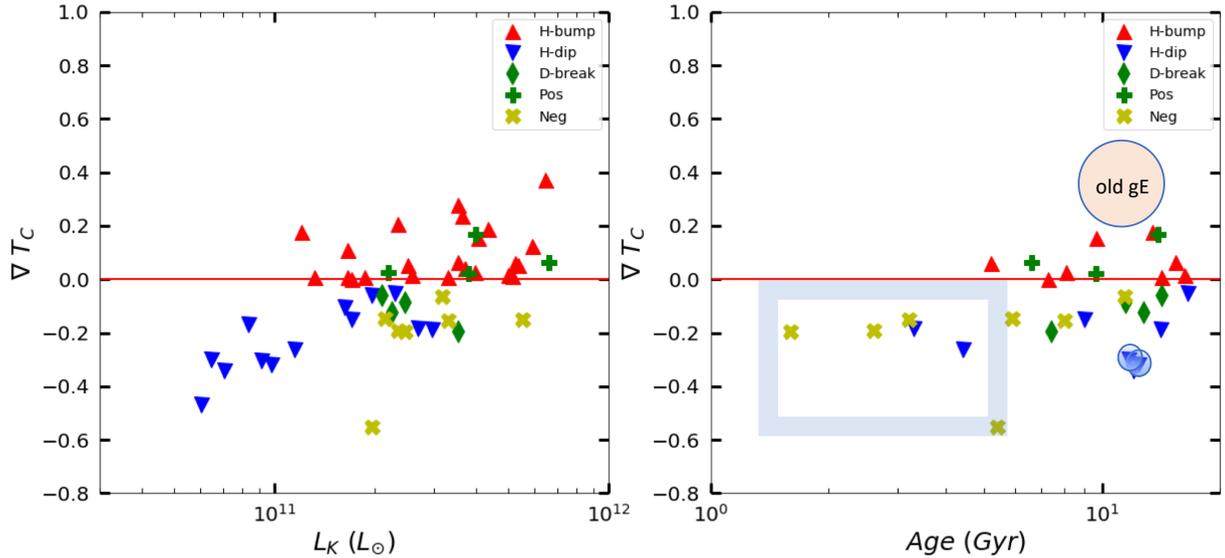

Figure 9. The core T-gradient ($\nabla T_{CORE}$) against (a) the K-band luminosity and (b) the stellar age. All symbols are the same as in Figure 5.

In the right panel of Figure 9, we plot $\nabla T_{CORE}$ against the stellar age. The correlation is weak, if any. The p-value is rather high (0.06-0.08). However, interestingly, all young galaxies with age < 5 Gyr (inside the box in Figure 9b) have a negative $\nabla T_{CORE}$, i.e., they have HC, possibly suggesting that the recent star formation may cause the HC phenomenon. The single stellar population (SSP) equivalent age of a few Gyr may actually mean that a smaller fraction (~5%) of the stellar mass is contained in an even younger population formed during the past < ~1 Gyr (e.g., Serra & Oosterloo 2010), because the SSP age is strongly biased toward the age of the youngest stars (see Serra & Trager 2007). The stellar feedback from these rejuvenated young stars at the central region may provide an additional heating source.

Table 4 Correlation Test between $\nabla T_{CORE}$ and another variable

```
------------------------------------------------------------------------------------
          number            linmix                    pearson       spearman
                  slope error  intercept error   rms  p-value   σ   p-value   σ
------------------------------------------------------------------------------------
All types

T_GAS     52      0.68  0.10    0.03   0.02    0.13   3.53e-09  5.9  3.07e-10  6.3   Fig.8
L_GAS     52      0.13  0.02   -5.36   0.85    0.14   2.60e-08  5.6  7.38e-10  6.2   Fig.8
L_K       52      0.35  0.07   -4.05   0.81    0.15   3.86e-06  4.6  2.42e-06  4.7   Fig.9
Age       30      0.01  0.01   -0.22   0.08    0.16   7.72e-02  1.8  5.54e-02  1.9   Fig.9
L_1.4     45      0.06  0.02   -1.66   0.47    0.16   9.66e-04  3.3  3.09e-04  3.6   Fig.10
L_AGN     52      0.10  0.03   -3.98   1.23    0.17   1.67e-03  3.1  3.77e-04  3.6   Fig.10
M_BH      28      0.09  0.05   -0.86   0.46    0.16   7.53e-02  1.8  2.75e-01  1.1   Fig.10
V_disp    50      0.64  0.28   -1.60   0.68    0.17   2.09e-02  2.3  1.32e-02  2.5   Fig.11
M_TOT     29      0.32  0.09   -3.90   1.06    0.15   5.59e-04  3.5  5.78e-04  3.4   Fig.11

With a hot core (H-dip + Neg + D-break)

T_GAS     25      0.31  0.17   -0.12   0.05    0.12   6.03e-02  1.9  2.02e-02  2.3   Fig.8
L_GAS     25      0.09  0.03   -3.80   1.29    0.11   5.36e-03  2.8  6.87e-04  3.4   Fig.8
L_K       25      0.18  0.08   -2.25   0.85    0.11   1.34e-02  2.5  1.75e-02  2.4   Fig.9
Age       19      0.01  0.01   -0.25   0.07    0.12   2.78e-01  1.1  1.62e-01  1.4   Fig.9
L_1.4     24      0.03  0.02   -1.13   0.55    0.12   7.46e-02  1.8  5.36e-02  1.9   Fig.10
L_AGN     25      0.03  0.04   -1.27   1.44    0.13   4.26e-01  0.8  5.61e-01  0.6   Fig.10
M_BH      16      0.04  0.05   -0.58   0.42    0.12   3.06e-01  1.0  2.54e-01  1.1   Fig.10
V_disp    25      0.28  0.30   -0.87   0.71    0.13   3.18e-01  1.0  3.41e-01  1.0   Fig.11
M_TOT     20      0.23  0.07   -2.90   0.89    0.11   2.54e-03  3.0  2.09e-03  3.1   Fig.11

No hot core (H-bump + Pos)

T_GAS     27      0.24  0.16    0.09   0.02    0.10   1.18e-01  1.6  1.89e-01  1.3   Fig.8
L_GAS     27      0.03  0.03   -1.20   1.10    0.10   2.11e-01  1.3  2.35e-01  1.2   Fig.8
L_K       27      0.11  0.10   -1.15   1.13    0.10   2.32e-01  1.2  1.07e-01  1.6   Fig.9
Age       11      0.00  0.01    0.04   0.09    0.07   6.05e-01  0.5  8.94e-01  0.1   Fig.9
L_1.4     21     -0.02  0.02    0.59   0.48    0.08   2.48e-01  1.2  4.69e-01  0.7   Fig.10
L_AGN     27      0.02  0.03   -0.77   1.24    0.10   4.51e-01  0.8  2.86e-01  1.1   Fig.10
M_BH      12      0.02  0.05   -0.07   0.48    0.07   7.22e-01  0.4  9.39e-01  0.1   Fig.10
V_disp    25     -0.34  0.24    0.91   0.59    0.08   1.39e-01  1.5  6.09e-01  0.5   Fig.11
M_TOT      9      0.03  0.22   -0.32   2.71    0.10   8.26e-01  0.2  3.81e-01  0.9   Fig.11
------------------------------------------------------------------------------------
```

- All tests were done between $\nabla T_{CORE}$ and the logarithm values in the first column, as in Figure 8 – 11.
- σ is the Gaussian-equivalent significance, assuming that the p-value is the double-side probability at $|x| > \sigma$ of the normal distribution. σ is marked as a boldface when p-value < 0.05.

It is still possible that recently formed stars may be embedded and hidden in some of the old galaxies. We further searched for the indirect evidence of recent star formation from the Atlas 3D (Cappellari et al. 2011) and Massive (Ma et al. 2014) surveys, both of which have extensively observed their sample ETGs in multi-wavelength facilities. Among our sample, 13 and 17 galaxies are in the Atlas 3D and Massive surveys, respectively. Since molecular clouds are the birthplace of stars, we looked for the CO detections and found one galaxy each from two surveys, NGC 4477 with $M_{H2} = 3 \times 10^7$ $M_\odot$ (Young et al. 2011) and NGC 383 with $1.7 \times 10^9$ $M_\odot$ (Davis et al. 2019). Similarly, one galaxy (NGC 4278) was detected in HI ($M_{HI} = 10^6$ $M_\odot$) in the core (Young et al. 2014). Note that we are considering the core HI detection, because the extended HI may not be directly related to the star formation. Interestingly, two galaxies (NGC 4278 and NGC 4477) have negative $\nabla T_{CORE}$ and one (NGC 383) has $\nabla T_{CORE} = 0.02$, only slightly above 0. In Figure 9b, the first two galaxies are marked by a large blue circle. NGC 383 does not have an age measurement. We also searched for the presence of dust, but none of the 13 galaxies in the Atlas sample have dust (Krajnovic et al. 2011).

We note that many galaxies with positive $\nabla T_{CORE}$ do not have age measurements, because the X-ray luminous galaxies (mostly hybrid-bump type) are often at larger distances than the X-ray faint galaxies due to the selection effect. Considering the tendency of the giant E (gE) galaxies that they are old systems, we mark the likely location of these old X-ray luminous galaxies in Figure 9, but this needs to be confirmed. O'Sullivan et al. (2017) also suggested that star formation is unlikely to be an enough heat source to significantly impact gas temperatures in dominant galaxies (BCGs) of groups.

In summary, $\nabla T_{CORE}$ is strongly correlated with $L_K$ (or $M_{STAR}$), indicating that additional inner heating is most effective in small systems. There is a possible hint that stellar feedback from recent star formation could be related to the internal heating mechanism of HC.

## 6.4 $\nabla T_{CORE}$ - AGN

To explore the AGN feedback in terms of its effect on the core temperature gradient, we first use the 1.4 GHz radio luminosity. The radio data were primarily taken from the collection of Brown et al. (2011) and supplemented by the NVSS of Condon et al. (1998). Because $L_{1.4GHz}$ is from the entire radio core and lobes, the radio emission of some galaxies with large radio lobes may not directly indicate the current AGN status. For example, for NGC 1316 (Fornax A) which is most luminous in our sample (the yellow x with the extreme $L_{1.4} \sim 10^{32}$ erg s$^{-1}$ Hz$^{-1}$ in Figure 10a), the core radio emission is only a small fraction of the total radio luminosity.

Another measure of the AGN strength is the X-ray luminosity at the galaxy center. In the CGA program (Kim et al. 2019a), we applied two-component spectral fitting with an APEC model for the soft component from the hot gas and a power-law model for the hard component from the point sources including LMXBs and AGN. We extract the hard component of the X-ray luminosity at the core. Because our sample galaxies do not host a strong AGN, i.e., they are mostly low-luminosity AGNs (LLAGN), we select the extraction radius at 3". This radius corresponds to the 96% (92%) encircled energy circle at 1.5 keV (4.5 keV) for an on-axis source so that the fraction of $L_{X,AGN}$ falling outside this radius is negligible. Since the X-ray luminosity from the LMXBs is also peaked at the galaxy center, we need to correct for it. We estimate the expected contribution from LMXBs by scaling from $L_K$ as in Kim & Fabbiano (2013). The central (3") region contains about ~5% (on average) of the total stellar light. For most case, the contribution from LMXBs are negligible, but $L_{X,AGN}$ may be reduced by a factor of two when it is below $10^{39}$ erg s$^{-1}$. In spite of the caveats of $L_{1.4\,GHz}$ and $L_{X,AGN}$, the two quantities are correlated with p-value ~ 5 x 10$^{-5}$ (see the bottom-left panel of Figure 10).

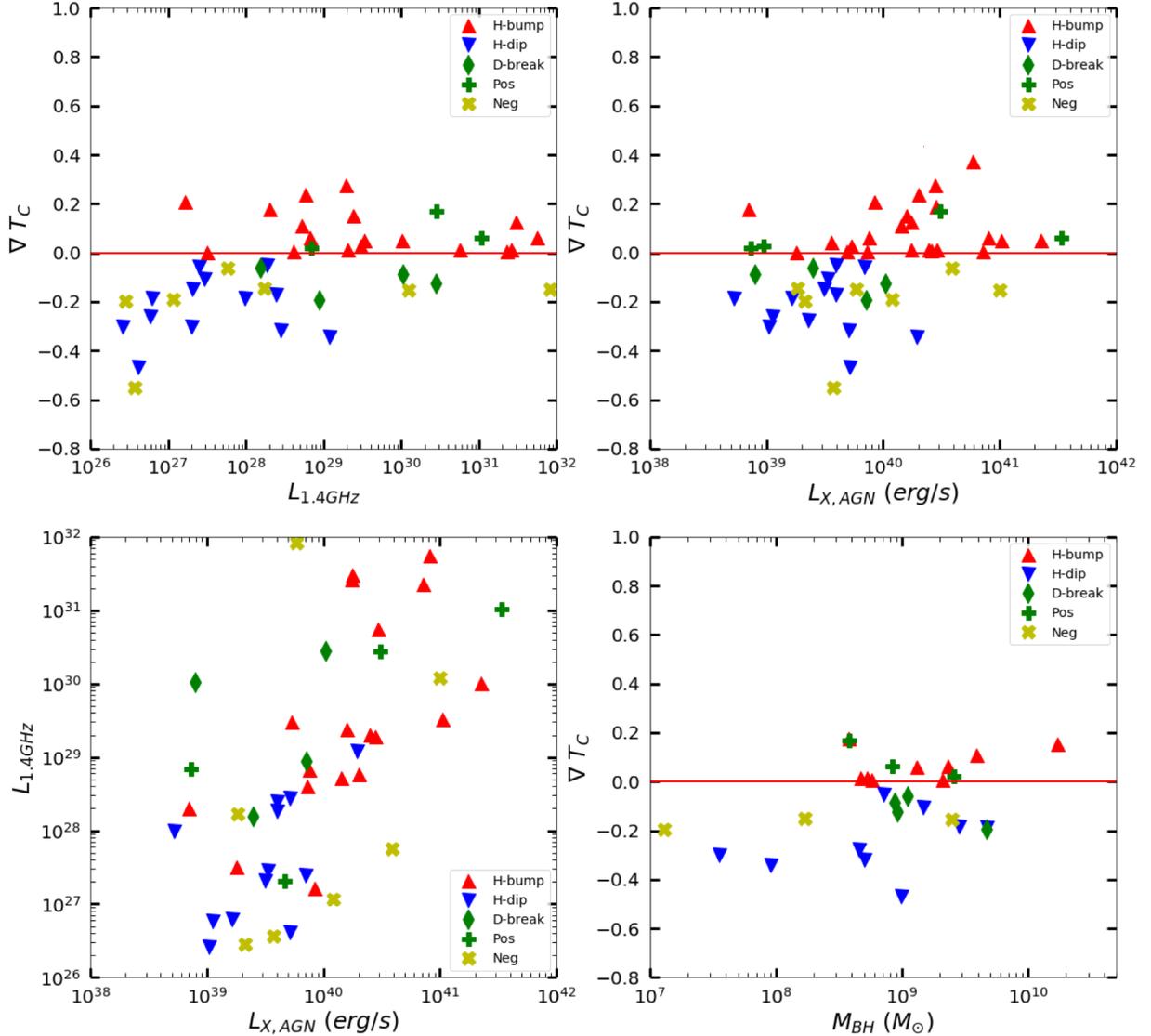

Figure 10. The core T-gradient (($\nabla T_{CORE}$) against (top left) $L_{1.4GHz}$, (top right) $L_{X,AGN}$, and (bottom right) $M_{BH}$. In the bottom left panel, $L_{1.4GHz}$ is compared with $L_{X,AGN}$. All symbols are the same as in Figure 5.

In the top row of Figure 10, we plot $\nabla T_{CORE}$ as a function of $L_{1.4GHz}$ (top left) and $L_{X,AGN}$ (top right). Both relations are at the ~3σ level for the entire sample, but almost no correlation for the subsamples with or without HC, except a weak (2σ) correlation between $\nabla T_{CORE}$ and $L_{1.4GHz}$ for the HC subsample. We also test with MBH (the mass of supermassive blackhole, see Table 1) in the bottom-right panel in Figure 10. Given the limited availability of the $M_{BH}$ data, we do not see a distinct correlation for the entire sample, nor the subsamples (see more in section 6.5).

The interpretation of the correlation strength is somewhat ambiguous. However, one obvious finding is that if any correlation exists, it is in a positive sense for the entire sample and the HC subsample. Interestingly, this means that the weaker the AGN strength is (in $L_{1.4GHz}$, $L_{X,AGN}$, and $M_{BH}$), the more pronounced the hot core is. Therefore, we found no evidence that HC is caused by the effect of the current AGN feedback. This may not completely rule out the AGN effect. Under the conceivable AGN cycle, AGN could have successfully heated its surroundings, shut off its own fuel supply, and become quiescent until cooling can build up a new reservoir of cold gas. However, even in this scenario, the hypothesis that the same galaxy evolves between HC and nHC is not supported, because the HC is preferentially found in small systems.

The weak negative correlation (p-value = 0.2 -0.5) between $\nabla T_{CORE}$ and $L_{1.4GHz}$ for the nHC subsample (the red and green points in the top left panel of Fig 10) may be interpreted as a symptom of the AGN feedback which plays a role in larger galaxies (e.g., in making the inner profile flatter for the hybrid-bump type), but the correlation is too weak and needs to be confirmed.

## 6.5 $\nabla T_{CORE}$ - $M_{TOT}$

To explore a possibility that the hot core may be produced by the gravitational heating as the hot gas cools and flows in, we further explore the relation with the central stellar velocity dispersion $\sigma_v$ (as a measure of the inner dynamical mass) and the total mass ($M_{TOT}$) of ETGs including stars and dark matter. We take $\sigma_v$ primarily from the Atlas3D and Massive surveys and supplement from the literature (see the references in Table 1). For $M_{TOT}$, we take the kinematically determined mass within 5 $Re$ from Alabi et al. (2017) and supplement with those scaled from the mass of the globular cluster system ($M_{GCS}$) assuming the near-linear relation between $M_{TOT}$ and $M_{GCS}$ (Kim et al. 2019b).

The correlation tests show ~$2.5\sigma$ (~$3.5\sigma$) correlations with $\sigma_v$ ($M_{TOT}$) for the entire sample (see Figure 11). Again, this is primarily driven by the fact that the bigger (smaller) system has a positive (negative) $\nabla T_{CORE}$. The correlations disappear in the subsamples, but the correlation between $\nabla T_{CORE}$ and $M_{TOT}$ remains strong (~$3\sigma$) for the HC subsample. This is in agreement with the correlations in Figure 8 between $\nabla T_{CORE}$ and the global hot gas properties ($L_{X,GAS}$ and $T_{GAS}$) which are known to be correlated with $M_{TOT}$ (e.g., KF13 and Kim et al. 2019). The meaningful correlations are all in a positive sense for the entire sample and the HC subsample. HC is more pronounced in galaxies with smaller $\sigma_v$ and/or smaller $M_{TOT}$. Our finding in this section, along with those in the previous section, i.e., only positive correlations with $M_{STAR}$ (represented by $L_K$) and $M_{BH}$, indirectly suggests that the gravitational heating of inflowing gas is not the source of HC. This is also supported by O'Sullivan et al. (2017), who found the inflow rates required for the central T rise in their group sample are unphysically high, a few solar masses per year.

Because the gravitational influence of the supermassive black hole may affect the hot gas temperature at the central region (see Humphrey et al. 2008 and 2009), it is of interest to look for any correlation with $M_{BH}$. However, $\nabla T_{CORE}$ is not correlated with $M_{BH}$, as seen in Figure 10 (bottom-right). It is still possible that we may not see this subtle effect because it works within r < a few x 100 pc (e.g., Pellegrini et al. 2012a). As in section 6.4, the only negative correlation between $\nabla T_{CORE}$ and $\sigma_v$ is for the nHC subsample, and it is very weak (p-value = 0.1 - 0.6). Again, this may be interpreted as a symptom of the gravitational heating in larger galaxies (e.g., in making the inner profile flatter for the hybrid-bump type), but this needs to be confirmed.

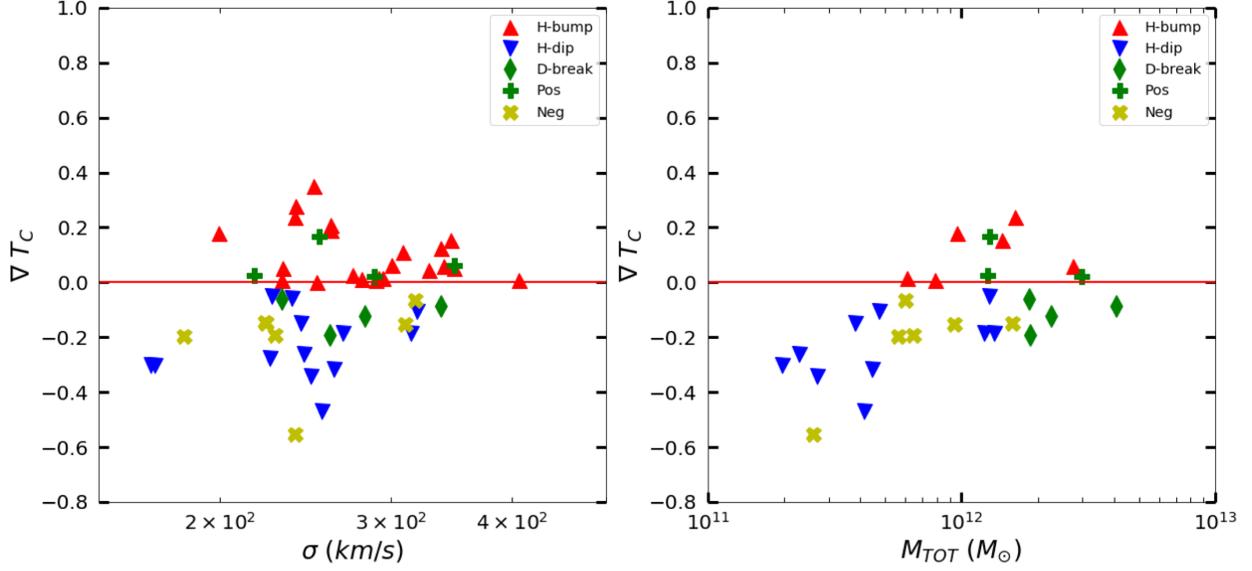

Figure 11. The inner T-gradient (($\nabla T_{CORE}$) against the stellar velocity dispersion, $\sigma_v$ and total mass, $M_{TOT}$.

## 7. CONCLUSIONS

Examining the CGA (Chandra Galaxy Atlas) data products of 60 early type galaxies (ETG) with extended hot halos, we found the following results.

- We identify six different types of temperature profiles: 26 hybrid-bump (rising at small radii and falling at large radii), 13 hybrid-dip (falling at small radii and rising at large radii), 8 negative (falling all the way), 6 positive (rising all the way), 4 double-break (falling at small radii, rising at intermediate radii, and falling again at large radii), and 3 irregular types.
- We find that the hot gaseous halos of the majority of early type galaxies in our sample can be explained with a universal temperature profile. For the hybrid-bump, hybrid-dip, and double-break types, the mean galacto-centric distance of the T peak is at $R_{MAX} = 35 \pm 25$ kpc (or ~0.04 $R_{VIR}$) and the mean distance of the T dip (or the inner break) is at $R_{MIN}$ (or $R_{BREAK}$) = 3 - 5 kpc (or ~0.006 $R_{VIR}$). The mean slope between $R_{MIN}$ ($R_{BREAK}$) and $R_{MAX}$ is $0.3 \pm 0.1$.
- The temperature gradient inside $R_{MIN}$ (or $R_{BREAK}$) varies widely, from negative, close to zero, to positive. The wide range of the core temperature gradient indicates the varying degree of additional heating at small radii. The hot core (HC) inside $R_{MIN}$ is most clearly visible for small galaxies. The nature of HC may be related to recent star formation, but we find no clear evidence that AGN feedback and gravitational heating play any significant role for HC.
- The positive type may also fit the universal profile when observational limits and selection effects are taken into consideration.
- The negative and irregular types are exceptions and require another explanation.

## ACKNOWLEDGEMENT

We have extracted archival data from the Chandra Data Archive, and the data analysis was supported by the CXC CIAO software and CALDB. We have used the NASA NED and ADS facilities. The computations in this paper were conducted on the Smithsonian High Performance Cluster (SI/HPC). This work was supported by the Chandra GO grants (AR5-16007X), by Smithsonian Competitive Grant Program for Science, by Smithsonian 2018 Scholarly Study Program, and by NASA contract NAS8-03060 (CXC). LT acknowledges support from the Southampton-Smithsonian exchange program.

## ACKNOWLEDGEMENT

We have extracted archival data from the Chandra Data Archive, and the data analysis was supported by the CXC CIAO software and CALDB. We have used the NASA NED and ADS facilities. The computations in this paper were conducted on the Smithsonian High Performance Cluster (SI/HPC). This work was supported by the Chandra GO grants (AR5-16007X), by Smithsonian Competitive Grant Program for Science, by Smithsonian 2018 Scholarly Study Program, and by NASA contract NAS8-03060 (CXC). LT acknowledges support from the Southampton-Smithsonian exchange program.

Table 1 Early type galaxy

| name | T | Dist | $R_e$ | $\log(L_K)$ | Age | $\log(\sigma)$ | $\log(L_{1.4})$ | $\log(M)$ BH | $\log(L_x)$ TOTAL | $\log(L_x)$ AGN | $\log(L_x)$ GAS | $T_{GAS}$ | $\nabla T_{CORE}$ | $T_{MAX}$ | $R_{MAX}$ | $T_{MIN}$ | $R_{MIN}$ | $R_{VIR}$ | Type |
|------|---|------|-------|-------------|-----|----------------|-----------------|--------------|-------------------|-----------------|-----------------|-----------|-------------------|-----------|-----------|-----------|-----------|-----------|------|
|      |   | Mpc  | kpc   | $L_{K\odot}$ | Gyr | km/s | erg/s/Hz | $M_\odot$ | erg/s | erg/s | erg/s | keV |  | keV | kpc | keV | kpc | Mpc |  |
| (1)  | (2) | (3) | (4) | (5) | (6) | (7) | (8) | (9) | (10) | (11) | (12) | (13) | (14) | (15) | (16) | (17) | (18) | (19) | (20) |
| I1262 | -5.0 | 130.2 | 7.7 | 11.4 |     | 2.37 | 29.5 |      | 41.0 | 43.2 | 1.30 | 0.05 | 1.72 | 67.9 | 0.93 | 8.5 | 0.92 | H-bump |
| I1459 | -5.0 | 29.2  | 5.2 | 11.5 | 8.0 | 2.49 | 30.1 | 12.0 | 9.4 | 41.0 | 40.6 | 0.48 | -0.15 |  |  |  |  | 0.56 | Neg |
| I1860 | -5.0 | 93.8  | 8.4 | 11.6 |     | 2.38 | 29.3 |      | 40.5 | 42.7 | 1.37 | 0.28 | 1.35 | 35.3 |  |  | 0.95 | H-bump |
| I4296 | -5.0 | 50.8  | 11.9 | 11.7 | 5.2 | 2.53 | 31.8 | 12.4 | 9.1 | 40.9 | 41.0 | 0.88 | 0.06 | 1.17 | 13.8 | 0.78 | 1.5 | 0.76 | H-bump |
| N0193 | -2.5 | 47.0  | 4.4 | 11.0 |     | 2.30 |      |      | 8.4 | 40.3 | 41.2 | 0.77 |      | 0.97 | 25.5 | 0.69 | 8.3 | 0.71 | H-bump |
| N0315 | -4.0 | 69.8  | 12.5 | 11.8 | 6.6 | 2.54 | 31.0 |      | 8.9 | 41.5 | 41.0 | 0.64 | 0.06 |  |  |  |  | 0.65 | Pos |
| N0383 | -3.0 | 63.4  | 6.3 | 11.5 |     | 2.46 | 31.4 |      | 8.8 | 40.9 | 41.3 | 1.35 | 0.01 | 1.73 | 64.7 | 0.81 | 2.0 | 0.94 | H-bump |
| N0499 | -2.5 | 54.5  | 4.3 | 11.3 |     | 2.38 | 27.4 |      |     | 39.8 | 42.2 | 0.70 | -0.06 |  |  | 0.72 | 9.9 | 0.68 | H-dip |
| N0507 | -2.0 | 63.8  | 12.9 | 11.6 | 8.1 | 2.44 | 29.5 |      |     | 39.7 | 42.9 | 1.32 | 0.03 | 1.34 | 69.6 |  |  | 0.93 | H-bump |
| N0533 | -5.0 | 76.9  | 16.2 | 11.7 |     | 2.45 | 29.3 |      |     | 40.4 | 42.0 | 0.98 | 0.01 | 1.41 | 32.5 | 0.81 | 4.4 | 0.80 | H-bump |
| N0720 | -5.0 | 27.7  | 4.8 | 11.3 | 5.4 | 2.38 | 26.6 | 11.4 |     | 39.6 | 40.7 | 0.54 | -0.55 |  |  |  |  | 0.60 | Neg |
| N0741 | -5.0 | 70.9  | 13.2 | 11.7 |     | 2.46 | 30.8 |      | 8.7 | 40.5 | 41.4 | 0.96 | 0.02 | 1.58 | 30.1 | 0.8 | 4.1 | 0.79 | H-bump |
| N1132 | -4.5 | 95.0  | 15.5 | 11.6 |     | 2.38 | 28.8 | 12.2 |     | 40.3 | 42.9 | 1.08 | 0.24 | 1.15 | 13.4 |  |  | 0.84 | H-bump |
| N1316 | -2.0 | 21.5  | 7.6 | 11.7 | 3.2 | 2.35 | 31.9 | 12.2 | 8.2 | 39.8 | 40.7 | 0.60 | -0.15 |  |  |  |  | 0.63 | Neg |
| N1332 | -3.0 | 22.9  | 3.0 | 11.2 |     | 2.50 | 27.5 | 11.7 | 9.2 | 39.5 | 40.4 | 0.41 | -0.11 |  |  | 0.56 | 3.1 | 0.52 | H-dip |
| N1380 | -2.0 | 17.6  | 3.2 | 11.1 | 4.4 | 2.39 | 26.8 | 11.4 |     | 39.1 | 40.0 | 0.30 | -0.26 |  |  | 0.28 | 2.1 | 0.44 | H-dip |
| N1387 | -3.0 | 20.3  | 3.5 | 11.0 |     | 2.23 | 27.3 | 11.3 |     | 36.7 | 40.6 | 0.41 | -0.30 |  |  | 0.35 | 4.2 | 0.52 | H-dip |
| N1395 | -5.0 | 24.1  | 5.4 | 11.3 | 7.6 | 2.40 | 26.9 |      |     | 40.0 | 40.4 | 0.65 |      | 0.93 | 10.8 |  |  | 0.65 | H-bump |
| N1399 | -5.0 | 19.9  | 4.7 | 11.4 | 11.5 | 2.53 | 30.0 | 12.6 | 8.9 | 38.9 | 41.5 | 1.21 | -0.09 | 1.44 | 22.6 | 0.93 | 0.6 | 0.89 | D-brk |
| N1400 | -3.0 | 26.4  | 2.9 | 11.0 | 15.0 | 2.41 | 27.3 | 11.3 |     | 39.7 | 40.0 | 0.55 |      |  |  |  |  | 0.60 | Pos |
| N1404 | -5.0 | 21.0  | 2.7 | 11.2 | 9.0 | 2.38 | 27.3 | 11.6 |     | 39.5 | 41.2 | 0.58 | -0.15 |  |  | 0.55 | 4.5 | 0.62 | H-dip |
| N1407 | -5.0 | 28.8  | 8.9 | 11.6 | 7.4 | 2.41 | 28.9 | 12.3 | 9.7 | 39.9 | 41.0 | 0.87 | -0.19 | 1.29 | 28.9 | 0.76 | 1.9 | 0.76 | D-brk |
| N1550 | -3.2 | 51.1  | 6.3 | 11.2 |     | 2.49 | 28.7 |      | 9.6 | 40.2 | 43.2 | 1.33 | 0.11 | 1.38 | 44.7 | 1.04 | 4.5 | 0.93 | H-bump |
| N1553 | -2.0 | 18.5  | 5.1 | 11.3 | 4.7 | 2.32 |      | 11.5 |     | 40.1 | 40.5 | 0.41 |      |  |  |  |  | 0.52 | Neg |
| N1600 | -5.0 | 57.4  | 13.5 | 11.6 | 9.7 | 2.54 | 29.4 | 12.2 | 10.2 | 40.2 | 41.5 | 1.43 | 0.15 | 1.48 | 46.4 | 1.03 | 4.3 | 0.97 | H-bump |
| N1700 | -5.0 | 44.3  | 3.9 | 11.4 | 2.6 | 2.36 | 27.2 | 11.8 |     | 40.1 | 41.0 | 0.43 | -0.19 |  |  |  |  | 0.53 | Neg |
| N2300 | -2.0 | 30.4  | 4.8 | 11.2 | 7.3 | 2.40 | 27.5 |      |     | 39.3 | 41.2 | 0.62 | 0.00 | 1.02 | 23.8 | 0.68 | 3.7 | 0.64 | H-bump |
| N2563 | -2.0 | 67.8  | 6.4 | 11.4 |     | 2.42 | 27.2 |      |     | 39.9 | 41.9 | 1.48 | 0.21 | 1.77 | 21.1 |  |  | 0.99 | H-bump |
| N3402 | -4.0 | 60.4  | 8.2 | 11.3 |     | 2.50 | 29.1 |      |     | 40.0 | 42.7 | 0.96 |      |  |  |  |  | 0.79 | Irr |
| N3923 | -5.0 | 22.9  | 5.8 | 11.4 | 3.3 | 2.43 | 26.8 | 12.1 | 9.5 | 39.2 | 40.6 | 0.45 | -0.18 |  |  | 0.45 | 3.2 | 0.54 | H-dip |

```
Table 1 - continued
```

| name | T | Dist | $R_e$ | $\log(L_K)$ | Age | $\log(\sigma)$ | $\log(L_{1.4})$ | $\log(M)$ TOTAL BH | | $\log(L_X)$ AGN GAS | | $T_{GAS}$ | $\nabla T_{CORE}$ | $T_{MAX}$ | $R_{MAX}$ | $T_{MIN}$ | $R_{MIN}$ | $R_{VIR}$ | Type |
|---|---|---|---|---|---|---|---|---|---|---|---|---|---|---|---|---|---|---|---|
| | | Mpc | kpc | $L_{K\odot}$ | Gyr | km/s | erg/s/Hz | $M_\odot$ | | erg/s | | keV | | keV | kpc | keV | kpc | Mpc | |
| (1) | (2) | (3) | (4) | (5) | (6) | (7) | (8) | (9) | (10) | (11) | (12) | (13) | (14) | (15) | (16) | (17) | (18) | (19) | (20) |
| N4104 | -2.0 | 120.0 | 20.0 | 11.8 | | | | | | 40.8 | 42.7 | 1.52 | 0.37 | 1.72 | 27.7 | | | 1.00 | H-bump |
| N4125 | -5.0 | 23.9 | 5.9 | 11.3 | 5.9 | 2.35 | 28.2 | | | 39.3 | 40.5 | 0.41 | -0.15 | | | | | 0.52 | Neg |
| N4261 | -5.0 | 31.6 | 6.9 | 11.4 | 16.3 | 2.47 | 31.4 | 11.8 | 8.7 | 40.2 | 40.9 | 0.76 | 0.02 | 1.35 | 10.2 | 0.77 | 2.7 | 0.71 | H-bump |
| N4278 | -5.0 | 16.1 | 2.6 | 10.9 | 12.0 | 2.40 | 29.1 | 11.4 | 8.0 | 40.3 | 39.4 | 0.30 | -0.34 | | | 0.31 | 1.9 | 0.44 | H-dip |
| N4291 | -5.0 | 26.2 | 2.0 | 10.8 | | 2.41 | 26.6 | 11.6 | 9.0 | 39.7 | 40.9 | 0.59 | -0.47 | | | 0.4 | 3.1 | 0.62 | H-dip |
| N4325 | 0.0 | 110.0 | 10.5 | 11.3 | | | | | | 40.4 | 43.1 | 1.00 | 0.01 | 1.08 | 51.2 | 0.83 | 6.8 | 0.81 | H-bump |
| N4342 | -3.0 | 16.5 | 0.5 | 10.1 | | 2.35 | | | 8.7 | 39.4 | 39.2 | 0.59 | -0.28 | | | 0.54 | 3.7 | 0.62 | H-dip |
| N4374 | -5.0 | 18.4 | 5.5 | 11.4 | 12.8 | 2.45 | 30.5 | 12.4 | 9.0 | 40.0 | 40.8 | 0.73 | -0.12 | 1.31 | 47.1 | 0.61 | 0.7 | 0.69 | D-brk |
| N4382 | -1.0 | 18.5 | 7.4 | 11.4 | 1.6 | 2.26 | 26.5 | 11.8 | 7.1 | 39.3 | 40.0 | 0.39 | -0.20 | | | | | 0.51 | Neg |
| N4406 | -5.0 | 17.1 | 10.3 | 11.3 | | 2.34 | | 12.1 | | 39.0 | 42.1 | 0.82 | 0.03 | | | | | 0.73 | Pos |
| N4438 | 0.0 | 18.0 | 5.0 | 10.9 | | 2.13 | 28.4 | | | 39.6 | 40.5 | 0.83 | -0.17 | | | 0.39 | 1.2 | 0.74 | H-dip |
| N4472 | -5.0 | 16.3 | 8.2 | 11.6 | 9.6 | 2.46 | 28.8 | 12.5 | 9.4 | 38.9 | 41.3 | 0.95 | 0.02 | | | | | 0.79 | Pos |
| N4477 | -2.0 | 16.5 | 3.5 | 10.8 | 11.7 | 2.23 | 26.4 | | 7.6 | 39.0 | 40.0 | 0.33 | -0.30 | | | 0.33 | 2.3 | 0.47 | H-dip |
| N4552 | -5.0 | 15.4 | 3.0 | 11.0 | 12.4 | 2.42 | 28.5 | 11.7 | 8.7 | 39.7 | 40.3 | 0.59 | -0.32 | | | 0.42 | 3.7 | 0.62 | H-dip |
| N4555 | -5.0 | 91.5 | 13.2 | 11.6 | | 2.52 | | | | 39.6 | 41.7 | 1.00 | 0.04 | 1.2 | 19.1 | 0.85 | 4.0 | 0.81 | H-bump |
| N4636 | -5.0 | 14.7 | 6.7 | 11.1 | 13.5 | 2.30 | 28.3 | 12.0 | 8.6 | 38.9 | 41.5 | 0.73 | 0.18 | 0.94 | 25.4 | 0.71 | 2.8 | 0.69 | H-bump |
| N4649 | -5.0 | 16.8 | 6.2 | 11.5 | 14.1 | 2.50 | 28.0 | 12.1 | 9.7 | 38.7 | 41.2 | 0.86 | -0.19 | | | 0.85 | 1.1 | 0.75 | H-dip |
| N4782 | -5.0 | 60.0 | 4.4 | 11.8 | | 2.53 | 31.5 | | | 40.2 | 41.4 | 1.15 | 0.13 | 1.58 | 55.2 | 0.83 | 8.5 | 0.87 | H-bump |
| N5044 | -5.0 | 31.2 | 3.9 | 11.2 | 14.2 | 2.37 | 28.6 | | | 39.9 | 42.4 | 0.91 | 0.01 | 1.33 | 53.9 | 0.93 | 5.4 | 0.77 | H-bump |
| N5129 | -5.0 | 103.0 | 14.3 | 11.6 | | 2.42 | | | | 40.5 | 42.6 | 0.81 | 0.19 | 1.04 | 19.5 | | | 0.73 | H-bump |
| N5171 | -3.0 | 100.0 | 12.4 | 11.3 | | | | | | 40.8 | 41.8 | 0.86 | | 1.27 | 15.7 | | | 0.75 | H-bump |
| N5813 | -5.0 | 32.2 | 8.3 | 11.4 | 16.6 | 2.35 | 28.3 | 12.1 | 8.9 | 39.6 | 41.9 | 0.70 | | | | | | 0.68 | Irr |
| N5846 | -5.0 | 24.9 | 7.2 | 11.3 | 14.2 | 2.37 | 28.2 | 12.3 | 9.0 | 39.4 | 41.7 | 0.72 | -0.06 | 1.08 | 29.5 | 0.71 | 2.2 | 0.69 | D-brk |
| N6338 | -2.0 | 123.0 | 17.1 | 11.7 | | 2.54 | 30.0 | | | 41.4 | 43.4 | 1.97 | 0.05 | 2.41 | 64.8 | 1.34 | 8.7 | 1.14 | H-bump |
| N6482 | -5.0 | 58.4 | 6.3 | 11.5 | 11.4 | 2.50 | 27.8 | 11.8 | | 40.6 | 42.2 | 0.74 | -0.06 | | | | | 0.70 | Neg |
| N6861 | -3.0 | 28.1 | 3.1 | 11.1 | | 2.61 | | 11.9 | 9.3 | 39.7 | 41.3 | 1.08 | 0.01 | 1.31 | 6.9 | 0.82 | 2.5 | 0.84 | H-bump |
| N6868 | -5.0 | 26.8 | 3.9 | 11.2 | 9.2 | 2.46 | | | | 39.9 | 41.3 | 0.69 | 0.02 | | | | | 0.67 | Pos |
| N7618 | -5.0 | 74.0 | 7.8 | 11.4 | | 2.47 | 29.4 | | | 40.5 | 42.3 | 0.80 | | | | | | 0.72 | Irr |
| N7619 | -5.0 | 53.0 | 8.8 | 11.6 | 15.4 | 2.48 | 28.8 | | 9.4 | 39.9 | 41.9 | 0.81 | 0.06 | 1.04 | 56.1 | | | 0.73 | H-bump |
| N7626 | -5.0 | 56.0 | 12.0 | 11.6 | 13.9 | 2.40 | 30.4 | 12.1 | 8.6 | 40.5 | 41.2 | 0.79 | 0.17 | | | | | 0.72 | Pos |

1. Galaxy name
2. Morphological type from RC3
3. Distance in Mpc taken mostly from SBF measurements, in order of preference, Tonry et al. (2001, ApJ, 546, 681), Jenson et al. (2003, ApJ, 583, 712), Cappellari et al. (2011, MNRAS, 413, 813), Lauer et al. (2007, ApJ, 664, 226), Tully et al. (2013, AJ, 146, 86). If not available, we take the distance from NED in http://ned.ipac.caltech.edu.
4. Effective radius ($R_e$) in kpc, taken from Atlas3D (Cappellari et al. 2013, MNRAS, 432, 1709), RC3 and 2MASS following the prescription in Cappellari et al. (2011, MNRAS, 413, 813), Lauer et al. (2007, ApJ, 664, 226), Blakeslee et al. (2001, MNRAS, 327, 1004), and NSA in http://www.nsatlas.org.
5. K-band luminosity. K mag is taken from 2MASS (via NED) and converted with $M_{K\odot}$ = 3.28 mag.
6. Stellar age assuming a single stellar population taken from, in order of preference, Thomas et al. (2005, ApJ, 621, 673), Terlevich & Forbes (2002, MNRAS, 330, 547), Trager et al. (2000, AJ, 120, 165), Kuntschner et al. (2010, MNRAS, 408, 97), Annibali et al. (2010, AA, 519, A40), Denicolo et al. (2005, MNRAS, 356, 1440), S'anchez-Bl'azquez et al. (2006, A&A, 457, 809).
7. Stellar velocity dispersion taken from Thomas et al. (2005, ApJ, 621, 673), Terlevich & Forbes (2002, MNRAS, 330, 547), Blakeslee et al. (2001 MNRAS, 327, 1004), Prugniel (1996, A&A, 309, 749), Gultekin et al. (2009, ApJ, 698, 198), Hyperleda in http://leda.univ-lyon1.fr
8. Radio luminosity in erg s$^{-1}$ Hz$^{-1}$ at 1.4 GHz, primarily taken from the collection of Brown et al. (2011, ApJ, 731L, 41) and supplemented by the NVSS by Condon et al. (1998, AJ, 115, 1693).
9. Total mass inside 5$R_e$, primarily taken from the GC kinematics by Alabi et al. (2017, MNRAS, 468, 3949) and supplemented by scaling from the mass of GC system by Kim et al. (2019b).
10. Mass of supermassive black hole taken from Kormendy, J. and Ho, L.C. (2013, ARAA, 51, 511), Gaspari et al. (2019, arXiv/1904.10972), Saglia et al. (2016, ApJ, 818, 4)
11. X-ray luminosity in 0.3 – 8 keV of the central source (see section 4)
12. X-ray luminosity in 0.3 – 8 keV of the hot gas from Kim et al. (2019a). For galaxies with extended halos, $L_{X,GAS}$ was measured from the entire hot halo with ROSAT or XMM-Newton data.
13. Temperature of the hot gas from Kim et al. (2019a).
14. Inner temperature gradient measured at 0.15$R_e$ (see section 6)
15. Temperature at the peak of the T profile bump
16. Galacto-centric distance at the peak of the T profile bump
17. Temperature at the bottom of the T profile dip for the hybrid dip and double break types, or at the inner break for the hybrid-bump type.
18. Galacto-centric distance at the bottom of the T profile dip for the hybrid dip and double break types, or at the inner break for the hybrid-bump type.
19. Virial radius in Mpc calculated by 0.81 x $T_{GAS}^{0.5}$ with $T_{GAS}$ from Col (13)
20. Temperature profile type determined in this work.

Appendix A

The temperature profiles of individual galaxies are grouped by the profile type. The red line is the parameterized 3D temperature model, and the blue line is the best-fit projected profile. The inner red vertical line indicates r = 3" where the AGN could affect the temperature measurement, and the outer red line indicates the maximum radius where the hot gas emission is reliably detected with an azimuthal coverage larger than 95%. The blue vertical line is at one effective radius.

1. Hybrid-bump

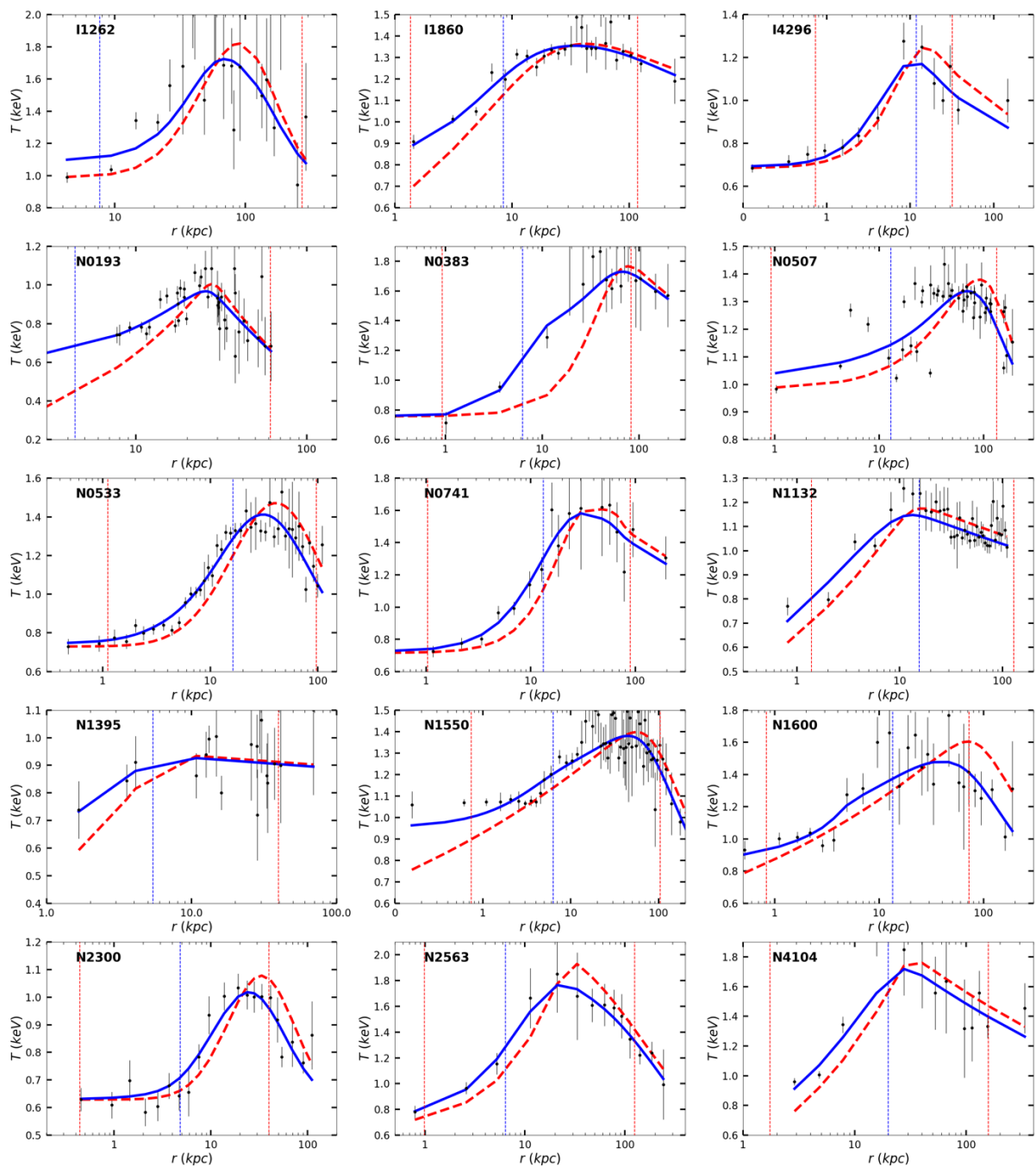

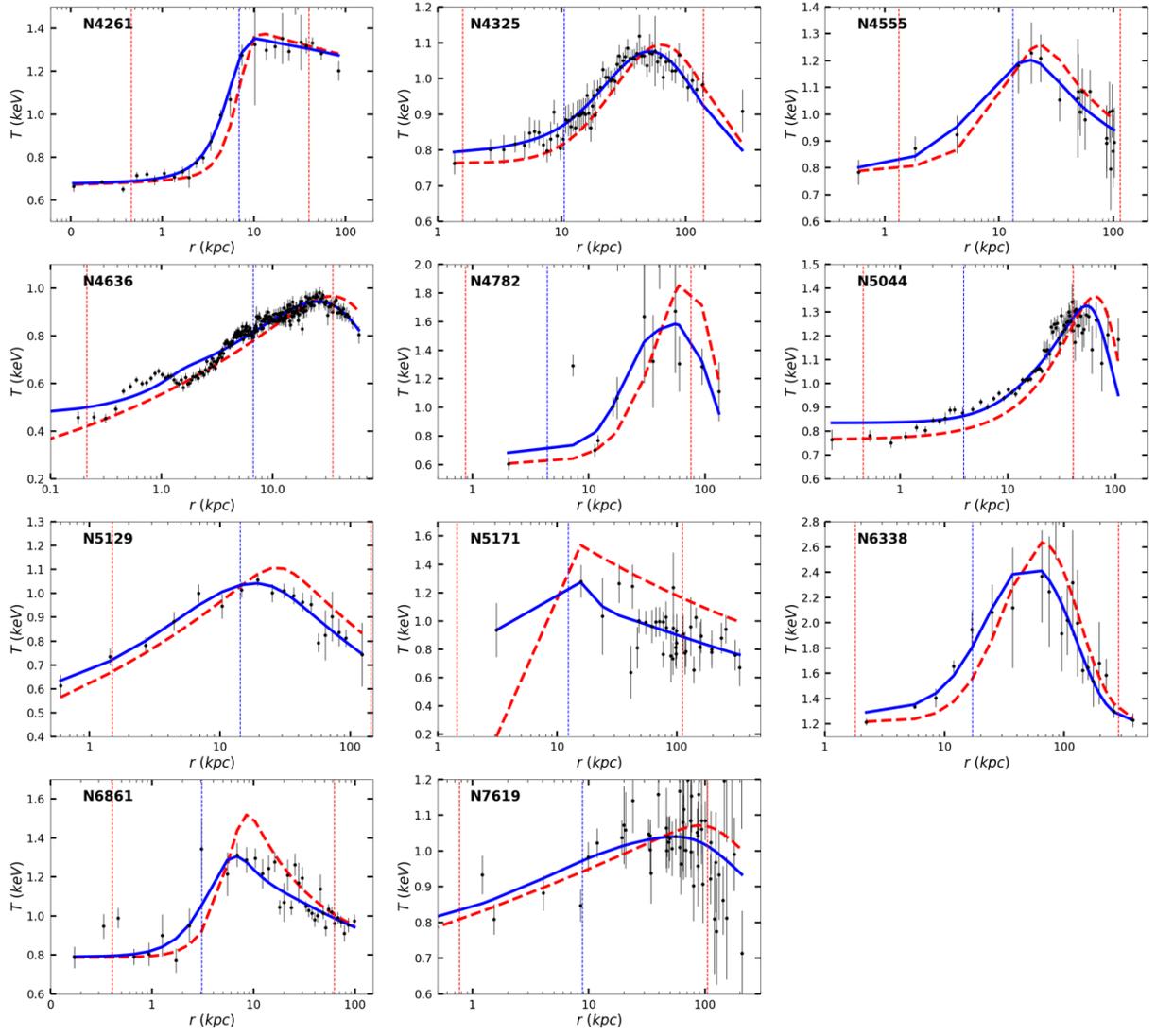

2. Hybrid-Dip

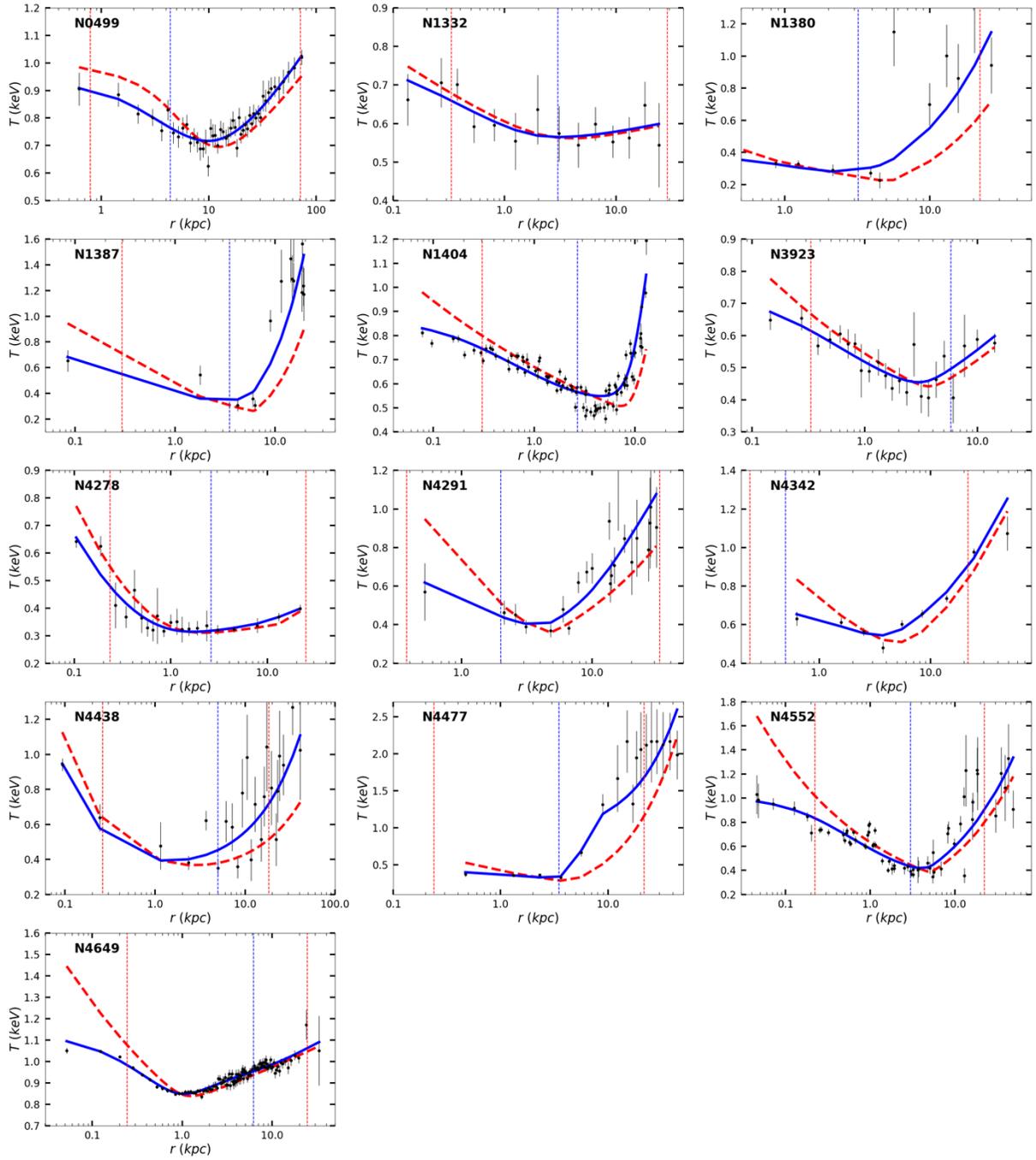

3. Double-Break

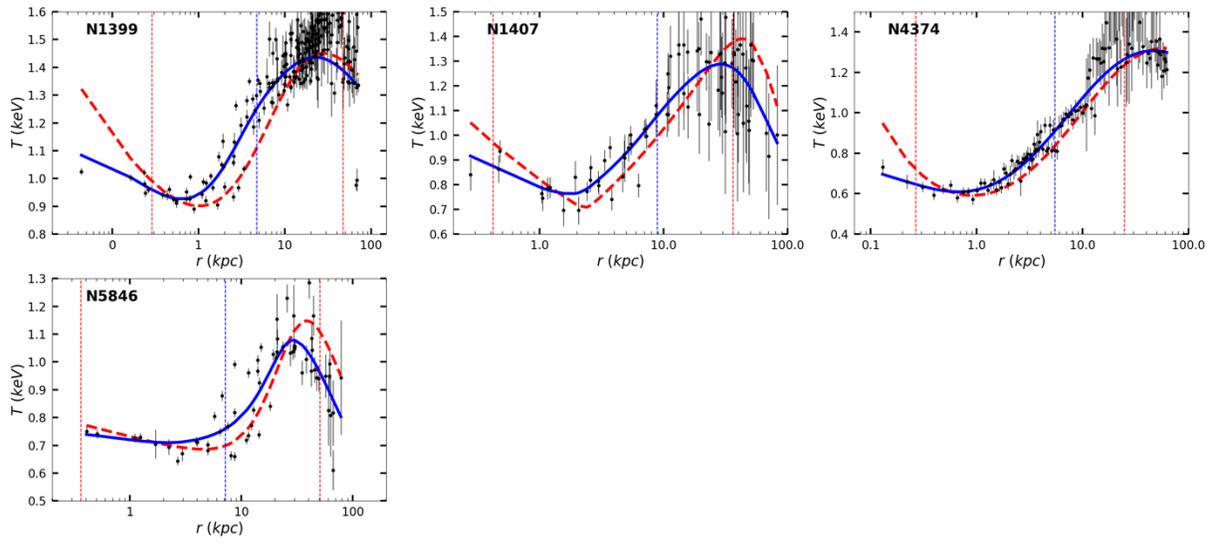

4. Positive

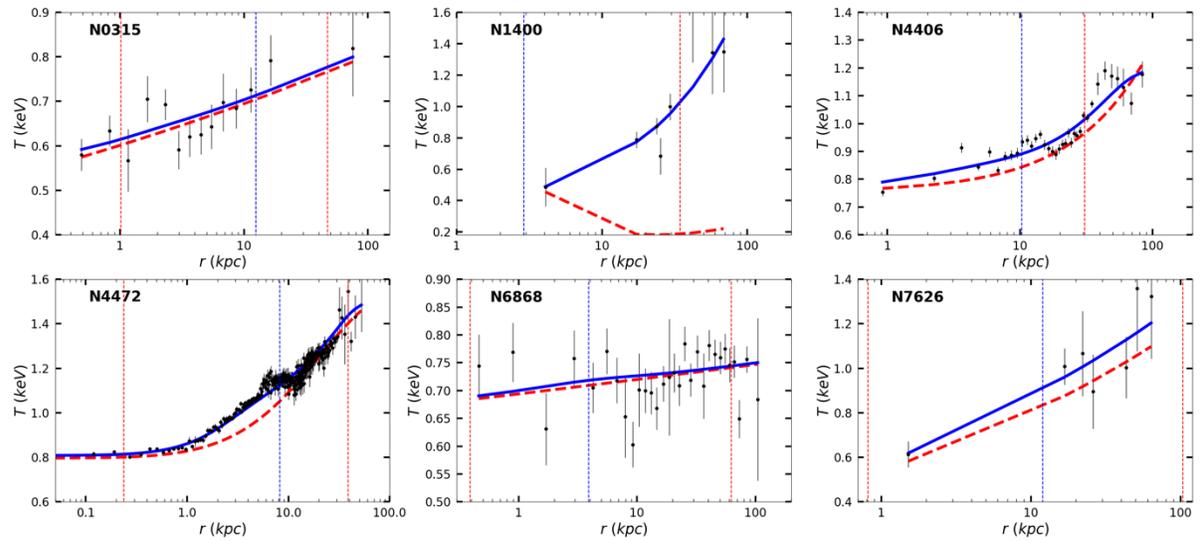

5. Negative

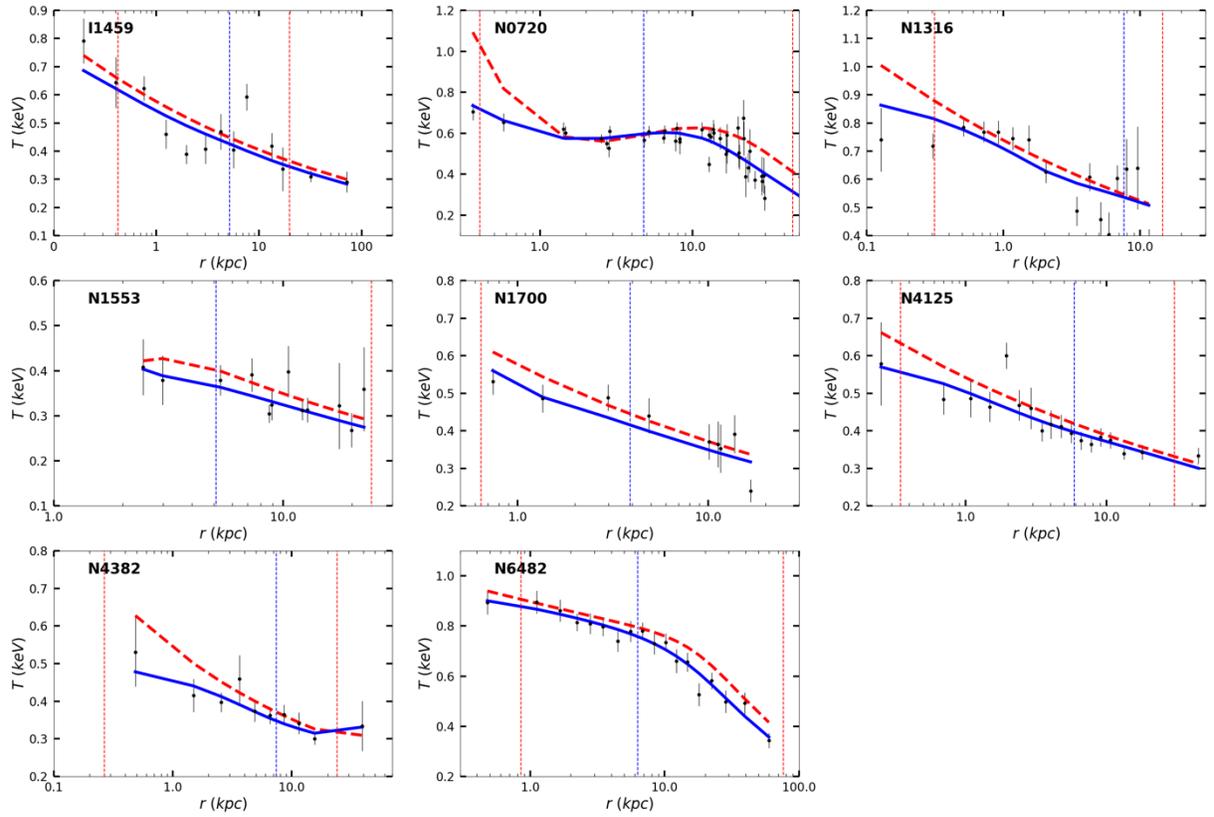

6. Irregular

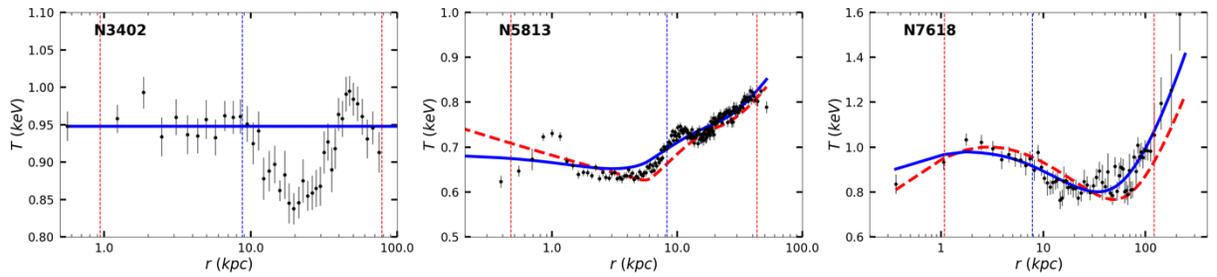